\documentclass[10pt,preprint2,usenatbib]{aastex}

\usepackage{amssymb}
\usepackage{graphicx}
\usepackage{multirow}
\usepackage{url}

\newcommand{\mum}{\ifmmode{\rm \mu m}\else{$\mu$m }\fi}

\newcommand{\chisq}{\ifmmode{\chi^{2} }\else{$\chi^2$}\fi}
\newcommand{\rchisq}{\ifmmode{\chi^{2} }\else{$\chi^2_\nu$}\fi}

\newcommand{\Hii}{H{\sc ii} }

\begin{document}


\title{The dustiest Post-Main sequence stars in the Magellanic Clouds}

\author{Olivia~C.~Jones\altaffilmark{1},
       {Margaret~Meixner\altaffilmark{1,2}},
       {Benjamin A.~Sargent\altaffilmark{3}},
       {Martha~L.~Boyer\altaffilmark{4,5}},  
       {Marta~Sewi{\l}o\altaffilmark{6,2}}, 
       {Sacha~Hony\altaffilmark{7}},
       {Julia~Roman-Duval\altaffilmark{1}}}
\email{ojones@stsci.edu}
\shorttitle{{\it Herschel} Evolved Stars in the MCs} 
\shortauthors{Jones et al.}

\altaffiltext{1}{Space Telescope Science Institute, 3700 San Martin Drive, Baltimore, MD, 21218, USA}
\altaffiltext{2}{The Johns Hopkins University, Department of Physics and Astronomy, 366 Bloomberg Center, 3400 N. Charles Street, Baltimore, MD 21218, USA}
\altaffiltext{3}{Center for Imaging Science and Laboratory for Multiwavelength Astrophysics, Rochester Institute of Technology, 54 Lomb Memorial Drive, Rochester, NY 14623, USA}
\altaffiltext{4}{Observational Cosmology Lab, Code 665, NASA Goddard Space Flight Center, Greenbelt, MD 20771, USA}
\altaffiltext{5}{Oak Ridge Associated Universities (ORAU), Oak Ridge, TN 37831, USA}
\altaffiltext{5}{Space Science Institute, 4750 Walnut St. Suite 205, Boulder, CO 80301}
\altaffiltext{7}{Institut f{\"u}r Theoretische Astrophysik, Zentrum f{\"u}r Astronomie, Universität Heidelberg, Albert-Ueberle-Str. 2, D-69120 Heidelberg, Germany}

\begin{abstract}
\noindent 
Using observations from the {\em Herschel} Inventory of The Agents of Galaxy Evolution (HERITAGE) survey of the Magellanic Clouds, we have found thirty five evolved stars and stellar end products that are bright in the far-infrared. These twenty eight (LMC) and seven (SMC) sources were selected from the 529 evolved star candidates in the HERITAGE far-infrared point source catalogs. Our source identification method is based on spectral confirmation, spectral energy distribution characteristics, careful examination of the multiwavelength images and includes constraints on the luminosity, resulting in a thoroughly vetted list of evolved stars. These sources span a wide range in luminosity and hence initial mass. We found thirteen low- to intermediate mass evolved stars, including asymptotic giant branch (AGB) stars, post-AGB stars, planetary nebulae and a symbiotic star. We also identify ten high mass stars, including four of the fifteen known B[e] stars in the Magellanic Clouds, three extreme red supergiants which are highly enshrouded by dust, a Luminous Blue Variable, a Wolf-Rayet star and two supernova remnants. Further, we report the detection of nine probable evolved objects which were previously undescribed in the literature. These sources are likely to be among the dustiest evolved objects in the Magellanic Clouds. The {\em Herschel} emission may either be due to dust produced by the evolved star or it may arise from swept-up ISM material.
\end{abstract}


\section{Introduction}
\label{sec:intro}

Evolved stars are important drivers of the chemical enrichment of the interstellar medium (ISM) of galaxies. 
In our own Galaxy dust enrichment of the ISM is dominated by the dusty winds from low- to intermediate mass stars (1--8 $M_{\odot}$) during the asymptotic giant branch (AGB) phase of evolution \citep{Gehrz1989}. At lower metallicities the relative contribution from high mass stars ($M > 8 M_{\odot}$) becomes important, and a significant period of dust production from these high-mass sources, via their stellar winds or directly in the supernova (SN) ejecta, is often invoked to explain the substantial dust abundances observed in high-redshift galaxies and quasars \citep{Dwek2007, Michalowski2010, Beelen2006, Bertoldi2002}.

Dust production at low-metallicities is only just beginning to be understood, due primarily to recent galaxy-wide surveys of nearby low metallicity systems such as the Magellanic Clouds (MC). 
The Large and Small Magellanic Clouds (LMC and SMC, respectively), with metallicities of 0.5 and 0.2 $Z_{\odot}$ \citep{Russell1992, Tchernyshyov2015}, are ideally suited to such investigations, due to their proximity (50 and 60 kpc; \citealt{Ngeow2008, Szewczyk2009}), favorable viewing angle (35$^{\circ}$ for the LMC; \citealt{vanderMarel2001}) and low average reddening along the line-of-sight ($E_{B-V}$ $\sim$ 0.08 and 0.04; \citealt{Schlegel1998}). 

The resolved stellar populations of the Magellanic Clouds have been well studied over a wide wavelength range from the optical to the far-infrared \citep[e.g.][]{ Zaritsky1997, Harris2004, Blum2006, Boyer2011}, as these surveys cover the full spatial extent of each galaxy, a complete picture of the mass-losing stellar populations in the Magellanic Clouds can been obtained. 
The dust-injection rate from low-mass stars in the Magellanic Clouds has been estimated by \cite{Srinivasan2009, Matsuura2009, Riebel2012, Boyer2012}, using mid-IR fluxes which are sensitive to warm ($\sim$ 100 K) dust. To compile a complete inventory of dust producers and fully characterise the stellar dust return to the ISM, the far-infrared (IR; 50--250 $\mu$m) and sub-millimeter (submm;  ($\lambda$ $\gtrsim $250 $\mu$m) emission from cold dust (T $<$ 50 K) envelopes associated with stars in an evolved stage of stellar evolution must also be constrained. 

Using data from the {\em Herschel} Inventory of the Agents of Galaxy Evolution in the Magellanic Clouds Open Time Key Project  (HERITAGE; \citealt{Meixner2013}), \cite{Seale2014} created a {\em Herschel} band-matched source catalog (BMC) for both the LMC and SMC by combining the individual HERITAGE catalogs of point sources at 100, 160, 250, 350, and 500~$\mu$m. They classify the sources in the BMC and find that the brightest HERITAGE sources are predominately young stellar objects (YSOs) while the faintest are background galaxies. Intriguingly, they provisionally identify a smaller population ($\sim$500) of very dusty stars, in the later stages of stellar evolution (e.g.~extreme AGB stars, post-AGB (PAGB) objects, planetary nebulae (PNe) and supernova remnants (SNR)), although the majority of these sources have also been identified as YSOs in other studies. 

In the Milky Way, the {\em Herschel} Key Program MESS (Mass loss of Evolved StarS; \citealt{Groenewegen2011}), has observed a representative sample of low- and intermediate-mass and high-mass post-main sequence objects, across a range of mass-loss rates. They find that extended structure in the far-IR is commonly seen around AGB stars and their remnants \citep{Kerschbaum2010, Cox2012, Maercker2012}.  
However, the presence of a significant number of evolved stars detected in the HERITAGE survey is unexpected, as it was predicted that the majority of these sources in the Magellanic Clouds would fall below the {\em Herschel} sensitivity limit \citep{Meixner2010AAS}. Only the most luminous, heavily dust enshrouded evolved stars were expected to have a detectable submm emission.  
The cold circumstellar envelopes of two massive evolved stars (R71 and IRAS 05280--6910) in the LMC have been detected with {\em Herschel} \citep{Boyer2010}. However, \cite{vanLoon2010} attribute this emission to swept-up ISM dust rather than from dust produced by the stars themselves.

The goal of this paper is to investigate the $\sim$500 evolved star candidates with a submm excess in the Magellanic Clouds identified using data from the {\em Herschel} HERITAGE-BMC \citep{Seale2014}. The paper is organised as follows: the observational data is summarized in Section~\ref{sec:data}. In Section~\ref{sec:sampleSelect} we describe the selection of post-main-sequence stars from the HERITAGE-BMC; in Section~\ref{sec:results} we present the results of this search and discuss the different post-main-sequence object classes contained in the catalog. Finally, Section~\ref{sec:conclusion} concludes this paper.

\section{Photometric Data}  
\label{sec:data}

Far-infrared (far-IR) and submm observations of the Magellanic Clouds were taken as part of the HERITAGE Key Project \citep{Meixner2013} using the Photodetector Array Camera and Spectrometer (PACS; \citealt{Poglitsch2010}) and the Spectral and Photometric Imaging Receiver (SPIRE; \citealt{Griffin2010}) on the {\it Herschel Space Observatory} \citep{Pilbratt2010}.  Photometric catalogs of point sources were created using {\sc starfinder} \citep{Diolaiti2000} which effectively removes diffuse emission to extract a point source flux. A full description of the HERITAGE observations and data reduction details can be found in \cite{Meixner2013}. 

The HERITAGE survey detected over 35000 unique sources in the LMC and 7500 in the SMC at 100, 160, 250, 350, and 500 $\mu$m. These sources were positionally cross-matched by \cite{Seale2014} to produce a HERITAGE Band-Matched catalog (BMC) for the LMC and SMC. This catalog was further matched to the {\em Spitzer} IRAC and MIPS catalogs from the Surveying the Agents of Galaxy Evolution (SAGE) {\em Spitzer} legacy surveys of the Large and Small Magellanic Clouds \citep{Meixner2006, Gordon2011}. Further details regarding the band-matching are given in \cite{Seale2014}. The final HERITAGE-BMC covers the 3.6 to 500 \mum range and contains over 42000 unique sources; these represent the dustiest populations of sources in the Magellanic Clouds.


In order to comprehensively classify the 500 sources in the HERITAGE photometric catalogs, and fully explore the dust production from point sources in the Magellanic Clouds, we have cross-matched the HERITAGE-BMCs with optical, near-IR and mid-IR point source catalogs, to provide photometry over a broader wavelength range (0.36 -- 500 $\mu$m).
The optical UBVI and near-IR JHK${_{\rm s}}$ photometry comes from the Magellanic Clouds Photometric Survey (MCPS; \citealt{Zaritsky1997}); the Infrared Survey Facility (IRSF) Magellanic Clouds Point Source Survey \citep{Kato2007}, the Two Micron All-Sky Survey (2MASS; \citealt{Skrutskie2006}), and the 2MASS 6X Deep Point Source Catalog (6X2MASS; \citealt{Cutri2004}). We also obtained supplementary mid-IR data from the {\em AKARI} LMC point source catalog  \citep{Ita2008, Kato2007} and the WISE all-sky catalog \citep{Wright2010}. 

We considered sources from the various catalogs to be a match if the source is within 5$''$ of the IRAC 3.6 \mum position in the BMC. In cases where there were multiple matches between catalogs we only considered the closest match. 

To complement the photometric data we searched the spectroscopic data from the {\em Spitzer} SAGE-Spec survey of the Magellanic Clouds \citep{Kemper2010, vanLoon2010} for all spectra taken with the InfraRed Spectrograph (IRS; 5.2-38 $\mu $m; Houck et al. 2004) and MIPS spectra (MIPS-SED; 52-97 $\mu $m) for matches to the HERITAGE-BMCs. We found 185 matches.  These sources have been spectroscopically classified by \cite{Woods2011, Ruffle2015} and Woods et al. (in prep).

\section{Refining the identification of post main sequence stars}
\label{sec:sampleSelect}

To identify the evolved stellar population in the Magellanic Clouds with a {\em Herschel} detection, we compiled a preliminary candidate list from sources which were identified as dusty objects in the late stages of stellar evolution including extreme and post-AGB stars, planetary nebulae, and supernova remnants. These sources were categorized by \cite{Seale2014} by matching the HERITAGE catalogs to other classification schemes and lists of previously identified objects in the literature. 

The HERITAGE BMCs contain 443 LMC and 51 SMC evolved star candidates and 45 LMC plus 9 SMC PN candidates \citep{Seale2014}. In total 455 of these Magellanic Cloud sources have also been assigned an alternate classification, making it difficult to judge the reliability of the source classifications in the initial list of evolved stars. For this reason we have to employ additional selection criteria to remove young stellar objects and background galaxies from the evolved star candidate list.

The following subsections describe our selection process, the methodology is generally listed in order of procedure however at some stages the selection criteria were assessed in parallel. For instance, an unambiguous identification can not be obtained solely upon the inspection of the source's spectral energy distribution (SED), thus the object's position and morphology was also considered simultaneously when assigning a likely classification.

\subsection{Spectroscopic confirmation}

The most reliable point-source classifications are obtained from spectroscopic observations. This alleviates most of the ambiguity that arises from photometric classification schemes due to overlaps in colour-magnitude space and thus provides robust source classifications for IR objects. We compare our initial list of evolved star candidates to the SAGE-Spec spectroscopic survey of the Magellanic Clouds \citep{Kemper2010}. Of the 529 unique sources in our list,  172 were found to have a spectroscopic counterpart. 

From IRS spectroscopy we find a total of seven evolved stars and four planetary nebulae.  Other less-common objects were also identified; these include one blue supergiant (BSG), one B[e] star, one Luminous Blue Variable (LBV), one Wolf-Rayet (WR) star, one symbiotic star and two sources of an unknown physical nature (\citealt{Woods2011, Ruffle2015}, Woods et al. in prep.).  Eight of these sources also have far-IR MIPS spectra covering the 52--97 \mum  wavelength range \citep{vanLoon2010, vanLoon2010b}. 
The remaining 154 objects with an IRS counterpart were confirmed spectroscopically as YSOs or \Hii regions \citep{Woods2011, Oliveira2013, Ruffle2015}. We remove these sources from the list of the evolved star candidates. The results of the spectral comparisons are summarised in Tables~\ref{tab:LMC_Fsample} and \ref{tab:SMC_Fsample}.

From the IRS spectra we have identified 18 post main sequence stars. Spectroscopic confirmation is still needed for the majority ($\sim60\%$) of the BMC evolved star candidates. For these  357 sources, we use several other complementary metrics  in parallel to  find evolved objects, improve upon the source identification and to exclude non-evolved contaminants from our sample.

\subsection{SED classification}

Photometrically it can be difficult to separate AGB stars from YSOs and young PNe from compact \Hii regions. 
The colour selection criteria for the far-IR sources from \cite{Boyer2011} and the candidate post-AGB stars from \cite{vanAarle2011} heavily overlap with the same regions as the YSO candidates in the colour magnitude diagrams (CMD). As such many of these sources have multiple classifications in the literature and we expect a high degree of contamination from YSOs in our initial evolved star sample selection. 

\begin{figure} 
\centering
\includegraphics[trim=1.0cm 0cm 0cm 0cm, width=3.5in]{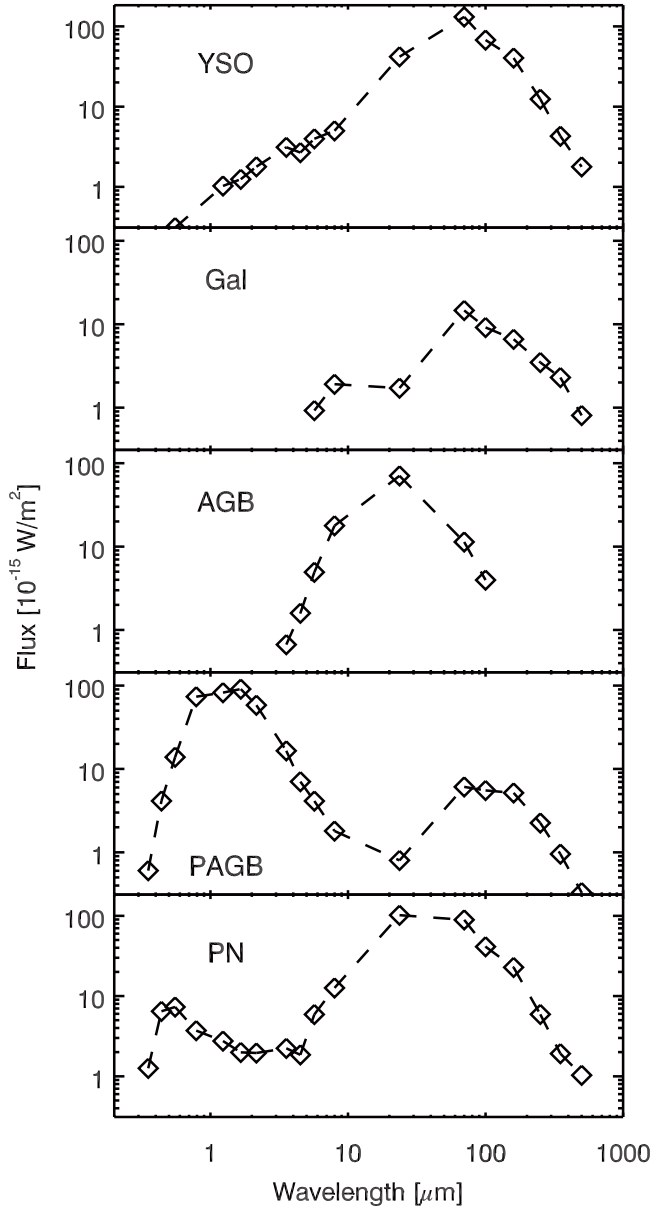}
 \caption[Example SEDs]{Example SEDs for the different groups of objects, including a YSO, a background galaxy, an AGB star with a large dust excess, a double-peaked SED indicative of a PAGB star, and a PN. The shape of the SED for YSOs differs markedly from the evolved stars.}
  \label{fig:egSEDs}
\end{figure}

To refine our selection further, the nature of each object with a far-IR detection was assessed by examining its SED from near-IR to far-IR wavelengths.
Figure~\ref{fig:egSEDs} shows some illustrative examples of typical SEDs for the different classes of objects.
Based on the shape of the SED with the flux plotted in log($\lambda$F$_{\lambda}$) against log($\lambda$) and more specifically on the peak of the energy distribution we can (tentatively) divide the sources in our list into four separate classes: (1) evolved star candidate, (2) probable YSO, (3) possible background galaxy, (4) a mismatch between the near- and far-IR data used in this study.

\subsubsection{Evolved star candidates}

The most extreme AGB stars can have red colours similar to YSOs, background galaxies, and PNe, and will also display a rising SED between 8 to 24 $\mu$m.  
We identify dust enshrouded AGB stars based on the peak of the SED; dusty AGB stars typically peak in the IRAC bands ($<$8 $\mu$m), while stars with little to no circumstellar dust peak in the near-IR. 
As the AGB star becomes more obscured the peak of the SED shifts to longer wavelengths, and the SED may peak between 8 and 24 \mum for the most extreme AGB stars. 
AGB stars detected with {\em Herschel} will fall in the latter category due to the cool dust envelopes which are likely providing additional extinction at the shorter wavelengths. 
In all cases the shape of the SED for AGB stars resembles that of a blackbody with temperature between 100 and 3000 K. 

 Red supergiants are more massive and are in general more luminous than AGB stars. As they exhibit a similar dust chemistry to oxygen-rich AGB stars, their SEDs are virtually indistinguishable. Ancillary information such as luminosity, pulsation period or membership of a young cluster is generally required to separate these classes (see Section~\ref{sec:Lum}).

Post-AGB stars tend to have a double-peaked SED, with one peak in the optical/near-IR (due to a hot central stellar component) and one in the mid-IR (due to a detached shell of circumstellar emission). This double-peaked morphology enables one to readily distinguish post-AGB stars from SEDs with a stellar component which has a large infrared excess. 

 The SEDs of other more massive evolved stars such as LBVs, WR stars and BSGs tend to be double-peaked, showing a stellar photosphere and a strong far-infrared excess. The classifications for these sources are taken from the literature \citep[e.g.][]{Bonanos2009, Bonanos2010}, and are not solely based on the SED.

\subsubsection{YSO candidates}

The SEDs of Stage 0/I YSOs (early to mid-stage YSOs) show a monotonic increase from the mid-IR to the far-IR, with the peak at $\sim$100 $\mu$m. Objects which peak in the far-IR are heavily obscured by cold dust, which absorbs the photospheric emission from the forming star, and re-radiates this energy at far-IR wavelengths.
As the YSO evolves it dissipates its surroundings and emission from warm dust begins to dominate the SED. For stars in the latest stages of formation, the peak emission is blue-ward of the far-IR, and these objects can be difficult to disentangle from other sources with a warm circumstellar dust continuum \citep{Oliveira2009}. Given the high likelihood of contamination from YSOs in our sample, additional care needs to be taken when inspecting the shape and peak of the energy distribution for these objects.  
However, objects which have an SED peak between 70 and 150 $\mu$m are likely YSOs \citep[e.g.][]{Sewilo2013}. 

To break the degeneracy between YSOs in the more advanced stage of formation and evolved stars which exhibit a significant IR excess, we created a mean YSO SED for each evolutionary stage based on the 154 YSOs with a spectroscopic counterpart \citep[see][for each YSO spectral class]{Woods2011}. These YSOs range from Stage 0/1--4 and cover a wide range in luminosity.
Using the YSO SED templates as a basis, we only retained those objects which peaked at wavelengths  $<$ 70\mum or had a significantly different profile to that of a YSO. We find 197 evolved star candidates closely mimic the SEDs of the previously identified YSOs.  After their removal, our sample contains 178 dusty evolved star candidates.

\subsubsection{PNe \& background galaxies}

The SEDs of PNe have a very similar appearance to those of YSOs, both in terms of shape and in the position of the peak; rising from the mid-IR and peaking in the far-IR.
Background galaxies also tend to have a rising SED in the mid-IR, and the peak of their SEDs tends to lie in the {\em Herschel} bands.
We can not resort to SEDs to distinguish between these classes.  We further consider potential confusion with background galaxies in Section~\ref{Image:gal}.

\subsubsection{Misclassification and mismatches}      

\begin{figure} 
\centering
\includegraphics[trim=1.0cm 0cm 0cm 0cm, width=3.5in]{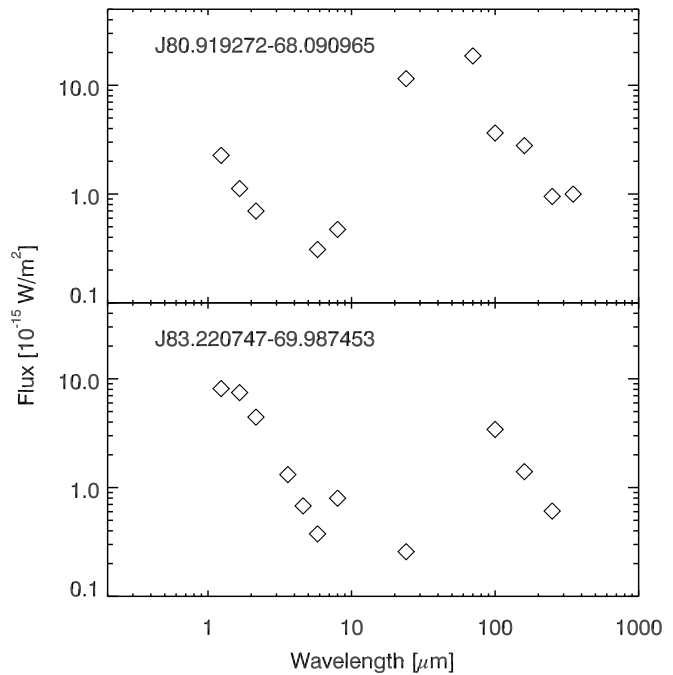}
 \caption[Example bad PAGB SEDs]{An example SED for a source in the BMC which is similar in appearance to that of a YSO (top) and for a source where there is a mismatch between the {\em Herschel} flux and the photometric data at shorter wavelengths (bottom). }
  \label{fig:egSEDs_PAGB}
\end{figure}

At low flux levels where a source is only detected in a few {\em Herschel} bands, the shape of the SED can be difficult to characterize. This can also be problematic in crowded fields, such as star-forming regions where there is a high probability of misidentification between photometric catalogs, due to the wide range in resolution between the surveys. We examine the SEDs for any large discontinuities between the photometry from different instruments and point source catalogs, indicative of a likely mismatch. 
An example of a mismatch between the photometric catalogs is shown in the bottom panel of Figure~\ref{fig:egSEDs_PAGB}. The SEDs for these sources are most likely a combination of a point source with warm circumstellar dust and spatially coincident starless dust clump which radiated in the far-IR. We eliminate  109 BMC evolved star candidates from the sample due to a discontinuity in the SED.

Low flux levels may also be problematic when comparing sources in the HERITAGE-BMC to objects previously identified in the literature, as this leads to higher positional uncertainties. This positional uncertainty in turn results in the potential misclassification of the BMC source as an evolved star. 
It is therefore vital that any source identified in the HERITAGE-BMC as an evolved star has a smooth SED with data points that appear consistent with each other and are associated with a point source. 

\subsubsection{SEDs of the final candidate list}

\begin{figure} 
\centering
\includegraphics[trim=2.5cm 0cm 0cm 0cm, width=3.5in]{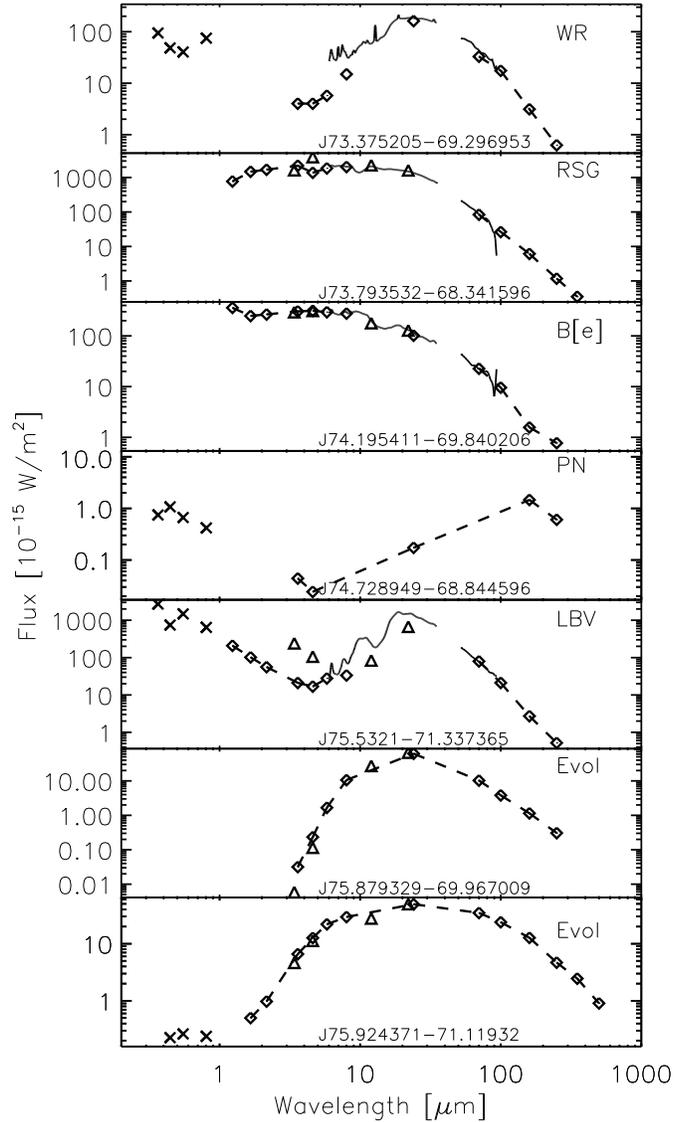}
 \caption[LMC SEDs]{SEDs and spectra for the post-main sequence sources in the LMC, ordered in terms of RA. Data from the HERITAGE-BMC and SAGE catalogs are shown as the dashed-diamond line, MCPS data as crosses, WISE data as triangles and AKARI data as squares.} 
  \label{fig:LMCSED_eg}
\end{figure}

\begin{figure} 
\centering
\includegraphics[trim=2.5cm 0cm 0cm 0cm, width=3.5in]{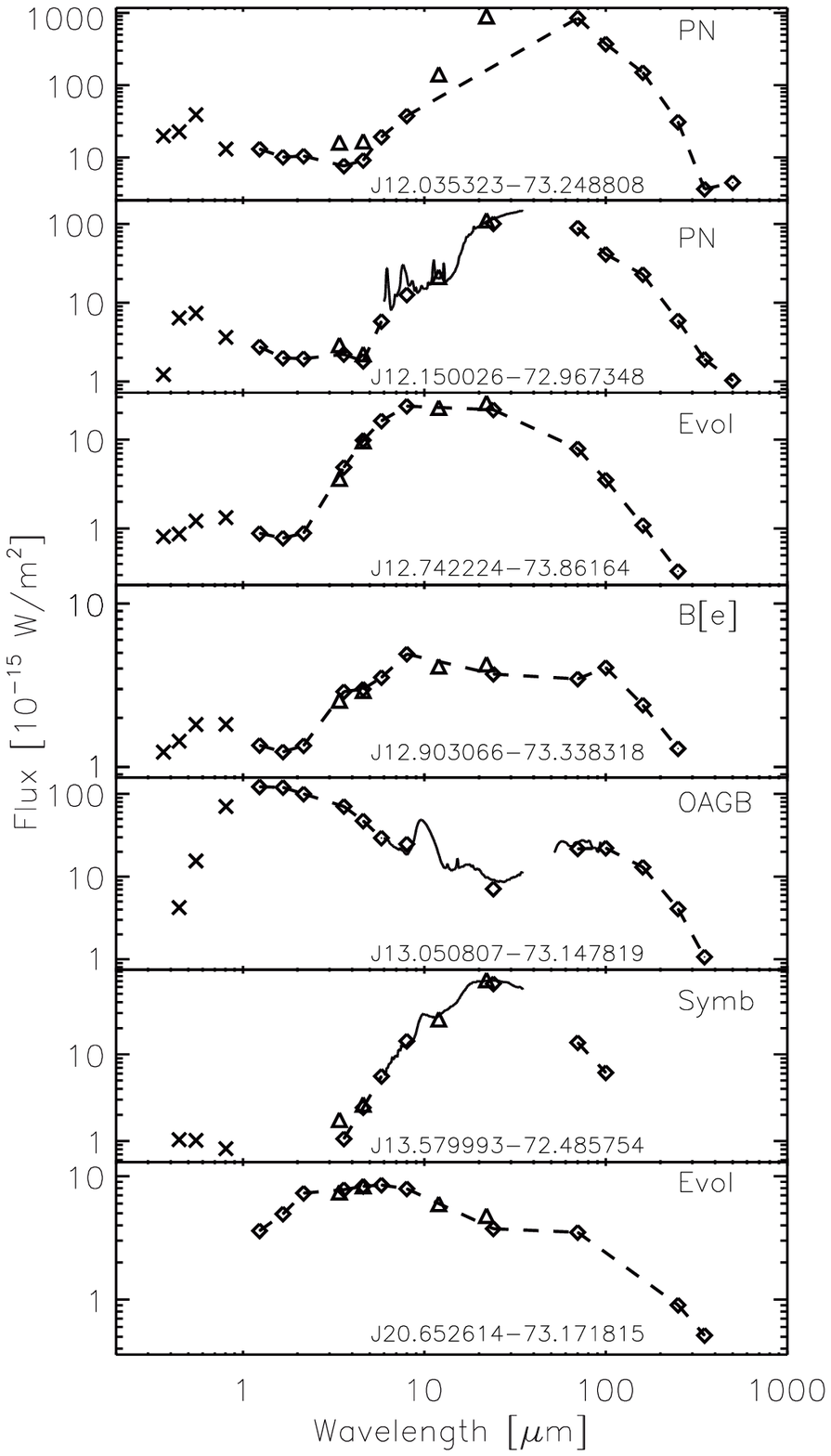}
 \caption[SMC SEDs]{As for Fig.~\ref{fig:LMCSED_eg}, but now for the SMC.}
  \label{fig:SMCSED}
\end{figure}

Figures~\ref{fig:LMCSED_eg} and \ref{fig:SMCSED} show the SEDs (as well as the available spectral data) for the final candidate list of post main-sequence objects in the MCs. In all instances a smooth transition is seen between the 2MASS, {\em Spitzer} and {\em Herschel} photometry, thus the shapes of the SEDs in the IR are well defined. At shorter wavelengths it is sometimes unclear if the MCPS data is associated with the source due to the difference in resolution, the high circumstellar extinction at optical wavelengths and intrinsic variability of the source. 

The SEDs show a range of distinguishing features due to the presence of warm and cold dust, which is dependent on the mass-loss history of the source. In all instances the SEDs are consistent with post main-sequence stars; the final classification is noted in each panel.

\begin{figure}
\centering
\includegraphics[trim=2.5cm 0cm 0cm 0cm, width=3.5in]{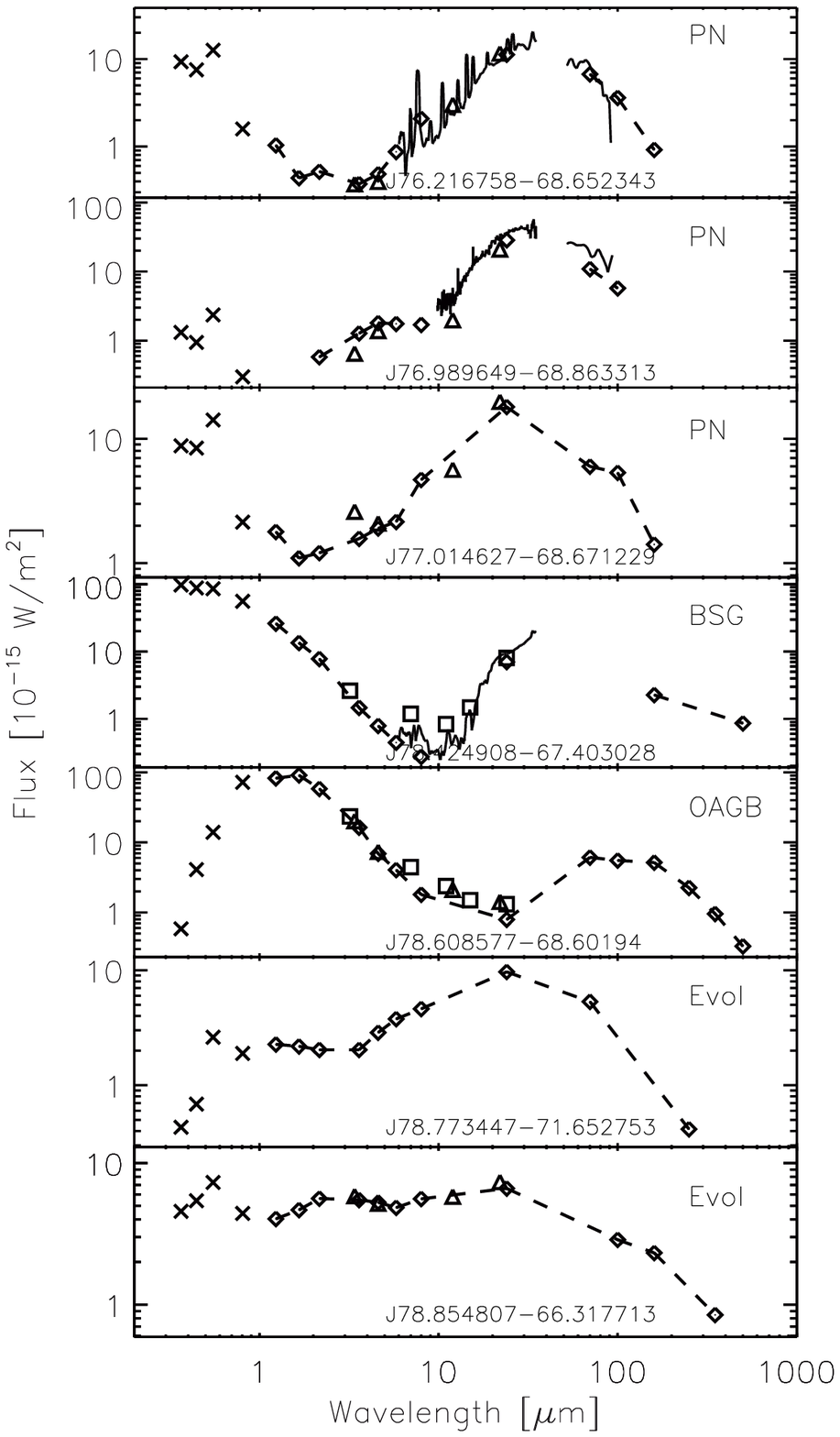}
\caption[]{Fig.~\ref{fig:LMCSED_eg}. Continued.}
\end{figure}

\begin{figure}
\centering
\includegraphics[trim=2.5cm 0cm 0cm 0cm, width=3.5in]{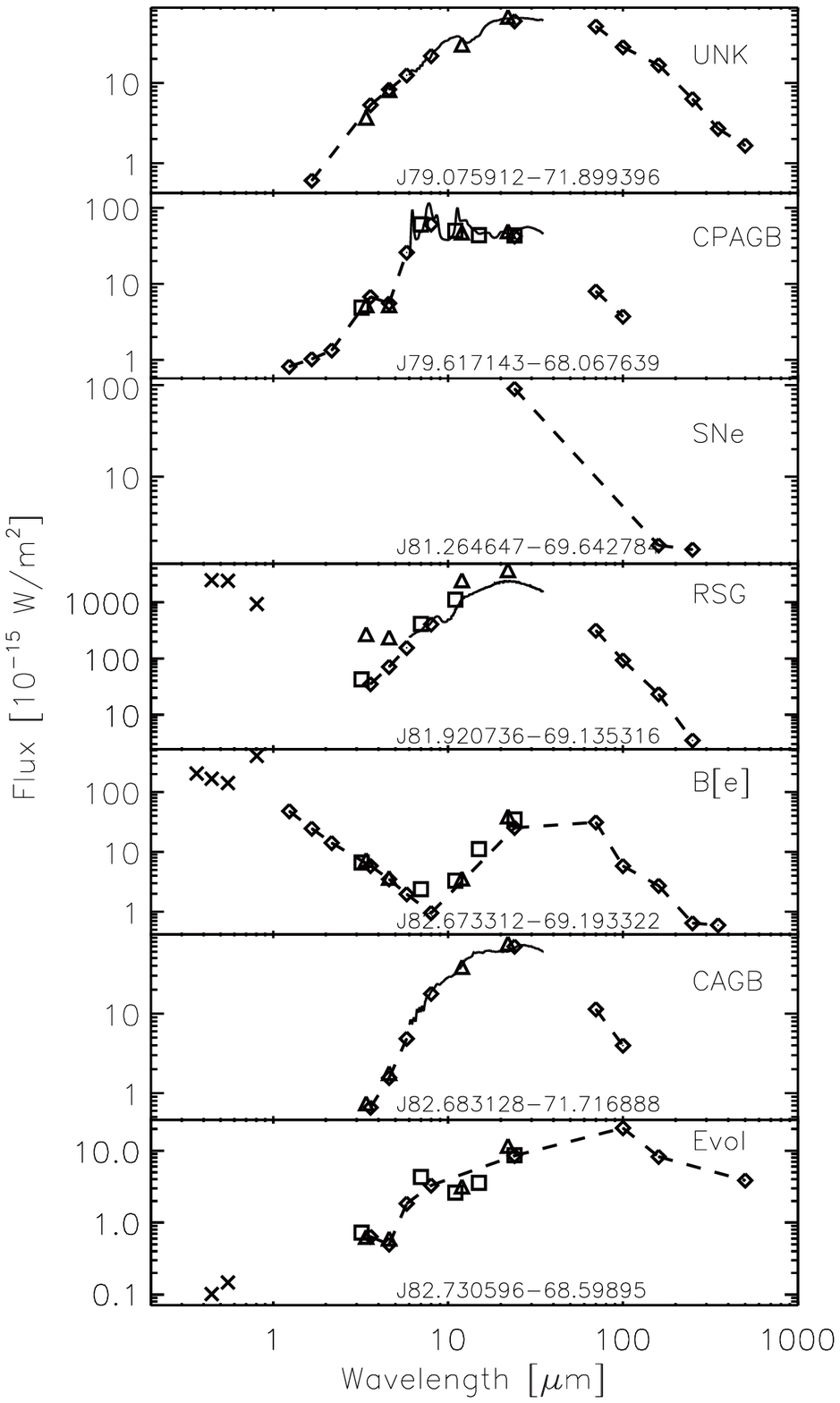}
\caption[]{Fig.~\ref{fig:LMCSED_eg}. Continued.}
\end{figure}

\begin{figure}
\centering
\includegraphics[trim=2.5cm 0cm 0cm 0cm, width=3.5in]{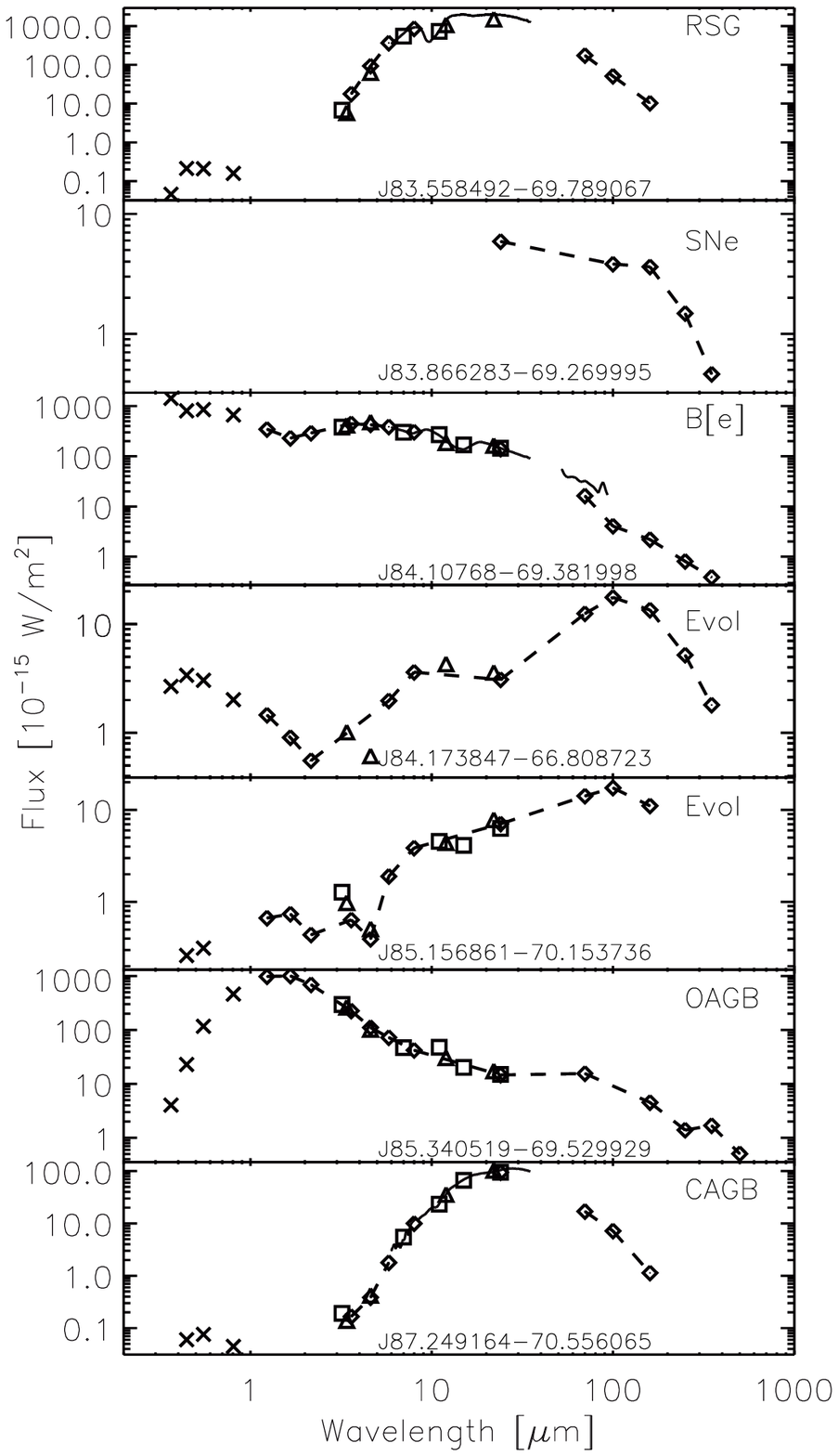}
\caption[]{Fig.~\ref{fig:LMCSED_eg}. Continued.}
\end{figure}

\subsection{Image inspection}

Using multiwavelength observations we assess the nature of each source in our initial sample. In several instances a source may only be identified in one or two {\em Herschel} bands and may be missing counterparts at other wavelengths.  Simutanious inspection of the images and the SEDs ensures that the BMC sources are real and not a composite of several nearby sources which contribute to the emission at different wavelengths, resulting from a mismatch between the {\em Herschel} and {\em Spitzer} catalogs. 

The PACS images are known to suffer from low level artifacts (in the form of striping and cross-hatching) \citep{Meixner2013} which may affect the quality and reliability of point sources extracted from the original images at 100 and 160 $\mu$m. Thus care needs to be taken when studying compact objects detected  in the PACS bands.
 
In an effort to obtain high-quality PACS images at 100 and 160 $\mu$m, the HERITAGE PACS data have been reprocessed with the Tamasis\footnote{{\url{http://pchanial.github.io/tamasis-pacs/}}} software (Chanal and Panuzzo, private communication) which mitigates thermal drifts in the PACS imaging using the redundancy and high pass filtering of the observations, effectively removing the artificial structures that compromised the original images. However, the Tamasis images cannot be used to extract photometry because the mapping technique and filtering makes the photometry potentially systematically inaccurate. Thus, we must inspect the refined PACS Tamasis images to distinguish between striping and point-sources in the HERITAGE BMC. We remove three spurious sources that are only detected with PACS and are coincident with the striped/hashed regions in the original images.

\subsubsection{Evolved stars vs.~YSOs}

To determine the likelihood that a given source is an evolved star we conduct a careful visual inspection of the IRAC, MIPS, PACS and SPIRE images.  Magellanic Cloud Emission Line Survey (MCELS; \citealt{MCELS1998}) H$\alpha$ images were also examined. Using DS9 \citep{Joye2003} we simultaneously view each evolved star candidate across multiple wavelengths, facilitating the inspection of the local stellar and interstellar environment. 

YSOs and dust clumps tend to be concentrated along bright ISM filaments \citep{Kim2010, Sewilo2013}; conversely, we expect there to be very few AGB stars associated with interstellar gas or dust structures. Evolved stars are generally isolated and inhabit regions of low column density.  Inspection of the images indicates that many of the HERITAGE-BMC evolved star candidates are embedded within bright knots of emission, located along large-scale dusty ISM structures, or are concentrated around molecular clouds; suggesting that these evolved star candidates are instead likely to be YSOs or massive stars. We remove sources located within sites of massive star formation from our list if their SED also resembles that of a YSO.

Visual inspection of the {\em Spitzer} images reveal a cluster-like morphology at the location of several of our evolved star candidates. This morphology is typical of small proto clusters, which appear as a single massive object in {\em Herschel}. 
The spatial resolution of the HERITAGE images is  $>2$ pc at the distance of the Magellanic Clouds, and blending of clustered sources is prevalent. The BMC  contains a GroupID which specifies group membership and is assigned to sources that have multiple matches in the BMC. As the majority of the sources in the catalog are not sorted into groups we use this criterion as an additional indication of evolutionary status and ensure that only isolated sources are retained.

\subsubsection{Background galaxies/AGN}
\label{Image:gal}

Some nearby galaxies are resolved with {\em Spitzer} IRAC and can be identified from their morphology; for one source in particular (HSOBMHERICC J85.622736-71.433709) we see a clear spiral pattern in the image. Other galaxies (e.g.~HSOBMHERICC J82.644586-68.104609) appear elongated and have an edge-on disc morphology, we remove  fifteen spatially resolved galaxies from the sample.

As a secondary diagnostic for background galaxies, \cite{Seale2014} identified distant galaxies in the BMC via their shape in the SPIRE images. Distant galaxies will be isolated from interstellar gas and dust structures, and appear unresolved and thus point-like at the {\em Herschel} wavelengths. Whereas, other HERITAGE source's, such as YSOs and dust clumps will appear slightly extended at SPIRE wavelengths.
For each source in the BMC, \cite{Seale2014} compares the sources 250 \mum full-width half-maximum (FWHM) to that of the point spread function (PSF). If the values are consistent to within 10\%, the source is defined to be point-like.  Twenty eight sources in our sample fulfill this criterion, although not all of them are necessarily background galaxies. Evolved stars without a detached shell will also appear point-like, thus we regard the shape in the far-IR as a rather weak constraint for separating background galaxies from AGB stars in the Magellanic Clouds,  and do not remove any object on this basis.

\subsection{Luminosity checks}
\label{sec:Lum}

To further refine our evolved star candidate list, we calculate the bolometric luminosity of the sources via a trapezoidal integration of the SED, from the optical to the far-IR \citep[e.g.][]{Woods2011, Jones2014, Ruffle2015}. All sources within each Magellanic Cloud are considered to be at approximately the same distance and we use a distance modulus of 18.5 to the LMC and 18.9 to the SMC \citep{Keller2006}. Where possible, we also compute  luminosities using the IRSF photometry rather than the 2MASS data and fold in the photometry from the WISE and AKARI surveys. 

\begin{figure} 
\centering
\includegraphics[width=3.5in]{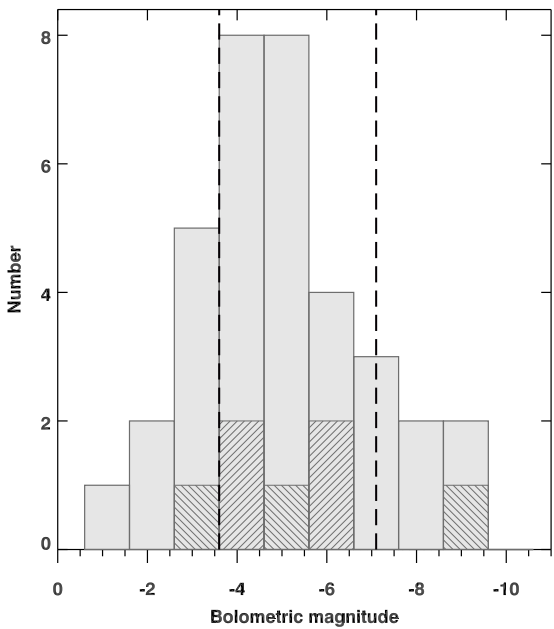}
 \caption[LuminosityFunction]{The luminosity distribution of the evolved star candidates in the LMC (solid grey) and SMC (black-striped). The dashed lines mark the expected luminosity range of AGB stars. } 
  \label{fig:Mbol}
\end{figure}

Figure~\ref{fig:Mbol} shows the absolute bolometric magnitudes distribution of the evolved star candidates in the LMC and SMC. The luminosites for the majority of the objects in our sample are consistent with evolved stars. Objects more luminous than the classical AGB luminosity limit of $M_{\rm bol}=-7.1$ \citep{Wood1983} may be RSGs or massive AGB stars undergoing hot bottom burning. At the low-luminosity end objects with $M_{\rm bol} > -3.9$ will be fainter than the tip of the Red Giant Branch (RGB) and are unlikely to be an evolved star, as stars on the RGB or early AGB do not experience significant dust formation. In general these limits are not an absolute division between the RGB, AGB and RSG classes, instead they provide useful guidance on the likely class.

\subsection{Literature classification}

To aid in the final classification, we performed a SIMBAD\footnote{{\url{http://simbad.u-strasbg.fr/simbad/}}} search on all of the sources to retrieve any other information pertinent to the classification.  We found that twelve of our sources are relatively unknown in the literature,  26 are identified as evolved stars or PNe with a high degree of confidence and 5 were classified as a YSO (see Section~\ref{sec:results}).  We removed known early type stars from our list if they were identified by means other than colour-magnitude cuts.

\subsection{YSO-fitter screen}

Finally we fit the evolved star candidates using the \cite{Robitaille2006, Robitaille2007} grid of YSO models, and visually examine the SED and model fits. We define well-fitted models as those whose $\chi^2$ per data point is less than ten. 

The YSO-fitter has been shown to be a good method of separating evolved stars from YSO candidates, since the YSO models do not span the high luminosities and relatively cool photospheres typical of evolved stars \citep{Whitney2008}. For other populations such as PNe the YSO fitter is less effective as a discriminant, as YSOs and PNe have similar SEDs and colours. For these sources a good fit does not inevitably imply that the source is a YSO. For the evolved star candidates 3 of the 38 sources were well fit by the YSO models; these sources are more likely to be true YSOs than a post-main sequence star, and thus they are removed from the sample.

\section{Results}
\label{sec:results}

\subsection{Classification Summary}

From the 529 evolved stars candidates provisionally identified in the HERITAGE BMCs, we find strong evidence for 35 post main sequence stars. Eighteen of these sources were confirmed spectroscopically while the remaining seventeen sources were selected using the stringent criteria described in Section~\ref{sec:sampleSelect}. In total 206 objects were rejected from our initial BMC sample as their SEDs and location suggests these are YSOs; a small number ($<17$) of these sources may potentially be evolved.

\begin{figure} 
\centering
\includegraphics[width=0.5\textwidth]{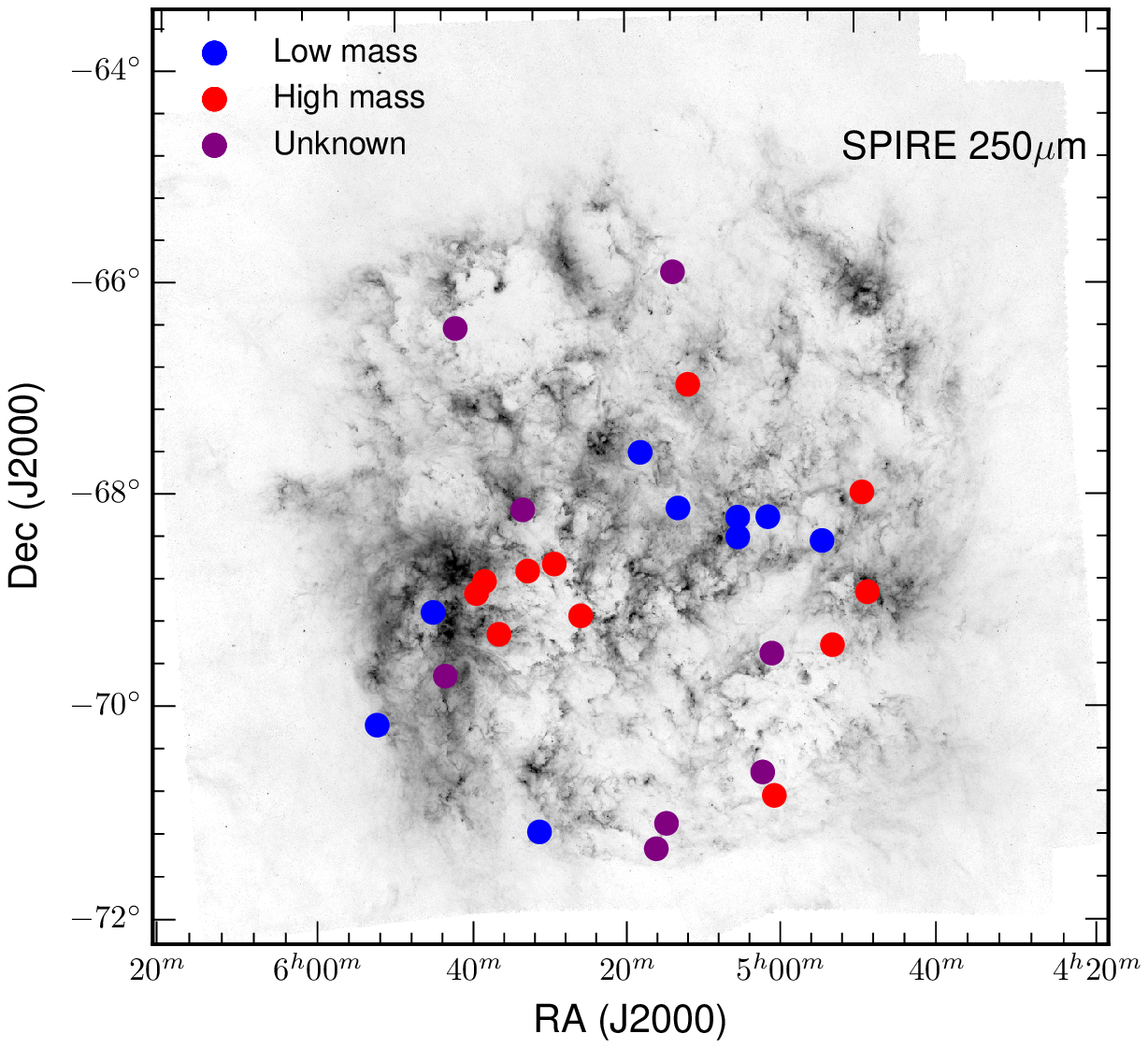}
\includegraphics[width=0.5\textwidth]{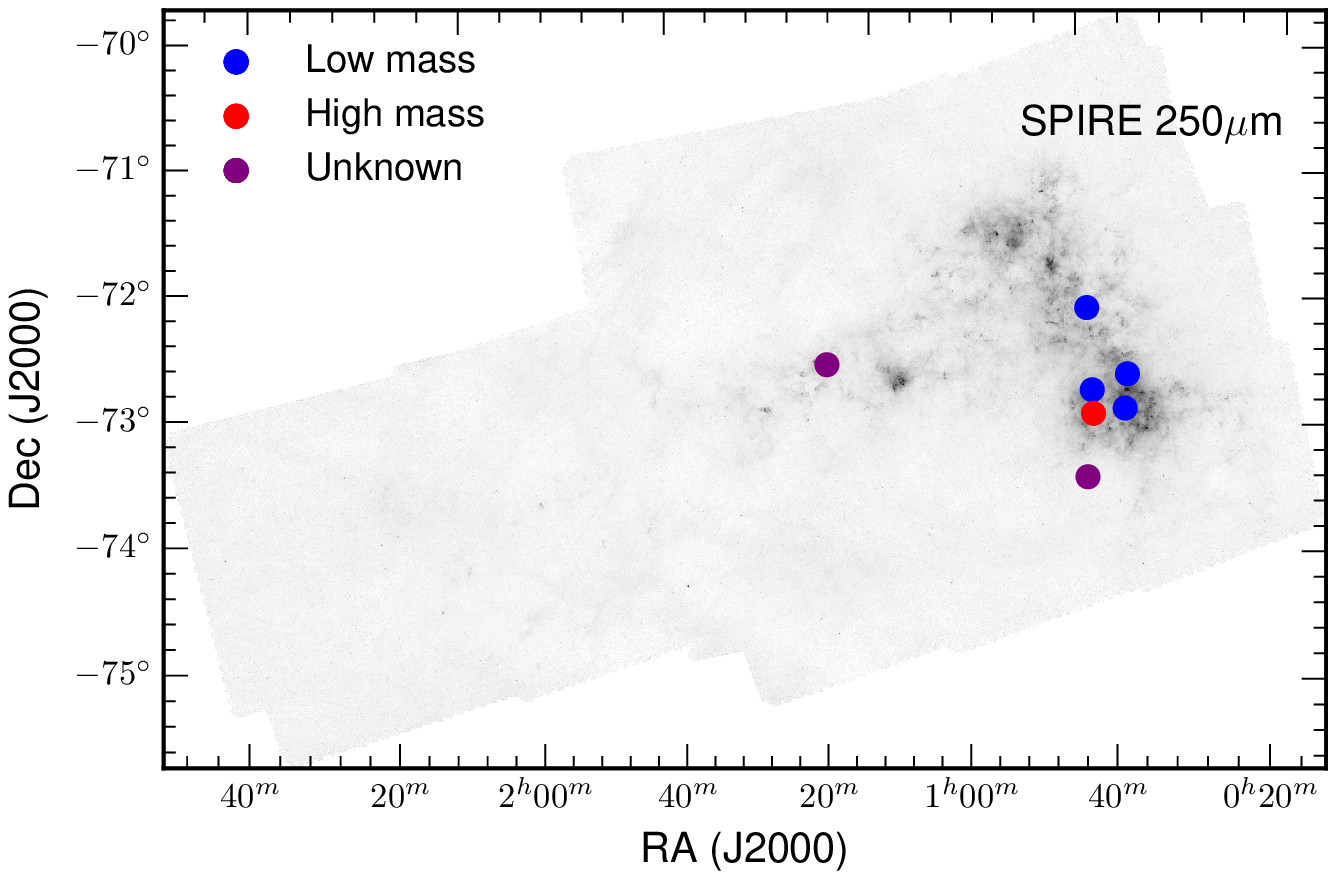}
\vspace{0.3cm}
 \caption[Sample Locations]{The BMC post-main sequence sources distributed on the sky, overlaid upon the HERITAGE SPIRE 250 \mum images of the LMC (top) and SMC (bottom).}
  \label{fig:BMCloc250}
\end{figure}

Tables~\ref{tab:LMC_Fsample} and \ref{tab:SMC_Fsample} list our final catalog of evolved stars, PNe and SNR with a {\em Herschel} detection in the LMC and SMC, respectively; their spatial distribution on the sky is displayed in Figure~\ref{fig:BMCloc250}, overlaid upon the HERITAGE 250 \mum images of the LMC and SMC. 

Figures~\ref{fig:LMCSED_eg} and \ref{fig:SMCSED} present the SEDs spanning optical to far-IR wavelengths of each object in the sample, and the photometry used to construct the SEDs is summarized in Table~\ref{tab:photTabsummary}. Finally, Table~\ref{tab:classtally} provides a tally of the various classes of post main sequence stars in the HERITAGE BMC. 
The following subsections discuss the individual objects in our final sample grouped according to their initial mass and evolutionary status. 

\begin{table*} 
\centering
\caption{Details of the final sample of evolved star candidates in the LMC. For each source, the BMC name, BMC coordinates, bolometric magnitudes ($M_{\rm bol}$), total luminosity, SAGE-Spec identification (SSID) and final classification is given (see Section~\ref{sec:results} for details).}
\label{tab:LMC_Fsample}
\centering
 \begin{tabular}{@{}lccccccc@{}}
   \hline
   \hline
BMC Source Name 	&     Alias                &	RA	&	Decl.	&	$M_{\rm bol}$	&	Luminosity 	&	SSID	&	Classification 	\\
                     	&                          &	(J2000)	& (J2000)	&	        	& ($L_{\odot}$)		&		&		\\
\hline 
J73.375205-69.296953	&       Brey 3a            &	04:53:30.0	&	$-$69:17:49	&	$-$5.9	&	17784	&	4060	&	WR	\\
J73.793532-68.341596	&       WOH G064           &	04:55:10.4	&	$-$68:20:30	&	$-$9.3	&	403277	&	4091	&	RSG 	\\
J74.195411-69.840206	&       R66                &	04:56:46.9	&	$-$69:50:25	&	$-$7.1	&	52908	&	4113	&	B[e]	\\
J74.728949-68.844596	&       RP 1805            &	04:58:54.9	&	$-$68:50:41	&	$-$1.5	&	331	&	24	&	PN	\\
J75.5321-71.337365	&       R71                &	05:02:07.0	&	$-$71:20:13	&	$-$7.1	&	55722	&	4162	&	LBV	\\
J75.879329-69.967009	&       \ldots             &	05:03:31.0	&	$-$69:58:01	&	$-$4.8	&	6668	&	\ldots	&	Possible evol.	\\
J75.924371-71.11932	&       \ldots             &	05:03:41.8	&	$-$71:07:10	&	$-$5.2	&	9874	&	\ldots	&	Possible evol.	\\
J76.216758-68.652343	&       SMP LMC 21         &	05:04:52.0	&	$-$68:39:08	&	$-$3.3	&	1715	&	4196	&	OPN	\\
J76.989649-68.863313	&       SMP LMC 28         &	05:07:57.5	&	$-$68:51:48	&	$-$4.1	&	3518	&	4209	&	OPN	\\
J77.014627-68.671229	&       SMP LMC 29         &	05:08:03.5	&	$-$68:40:16	&	$-$3.8	&	2516	&	\ldots	&	PN	\\
J78.424908-67.403028	&       BSDL 923           &	05:13:41.9	&	$-$67:24:11	&	$-$3.5	&	1913	&	69	&	BSG	\\
J78.608577-68.60194	&       \ldots             &	05:14:26.0	&	$-$68:36:07	&	$-$4.6	&	5374	&	\ldots	&	O-AGB/PAGB	\\
J78.773447-71.652753	&       \ldots             &	05:15:05.6	&	$-$71:39:10	&	$-$3.4	&	1776	&	\ldots	&	Possible evol.	\\
J78.854807-66.317713	&       \ldots             &	05:15:25.0	&	$-$66:19:04	&	$-$3.4	&	1825	&	\ldots	&	Possible evol.	\\
J79.075912-71.899396	&       IRAS 05170--7156   &	05:16:18.2	&	$-$71:53:58	&	$-$5.4	&	11702	&	78	&	UNK	\\
J79.617143-68.067639	&       IRAS 05185--6806   &	05:18:28.1	&	$-$68:04:04	&	$-$4.9	&	7201	&	4307	&	C-PAGB	\\
J81.264647-69.642784	&       N 132D             &	05:25:02.0	&	$-$69:38:39	&	\ldots	&	\ldots	&	\ldots	&	SNR	\\
J81.920736-69.135316	&       IRAS 05280--6910   &	05:27:46.2	&	$-$69:08:07	&	$-$7.8	&	107347	&	4454	&	RSG	\\
J82.673312-69.193322	&       [BE74] 328         &	05:30:41.5	&	$-$69:11:36	&	$-$4.7	&	6224	&	\ldots	&	B[e]	\\
J82.683128-71.716888	&       IRAS 05315--7145   &	05:30:43.9	&	$-$71:43:01	&	$-$5.0	&	7688	&	125	&	C-AGB	\\
J82.730596-68.59895	&       \ldots             &	05:30:55.3	&	$-$68:35:56	&	$-$4.2	&	3938	&	\ldots	&	Possible evol.	\\
J83.558492-69.789067	&       IRAS 05346--6949   &	05:34:14.0	&	$-$69:47:21	&	$-$7.7	&	91262	&	4555	&	RSG	\\
J83.866283-69.269995	&       SN 1987A           &	05:35:28.0	&	$-$69:16:11	&	\ldots	&	\ldots	&	\ldots	&	SNR	\\
J84.10768-69.381998	&       R126               &	05:36:25.8	&	$-$69:22:55	&	$-$7.3	&	64377	&	4601	&	B[e]	\\
J84.173847-66.808723	&       \ldots             &	05:36:42.0	&	$-$66:48:31	&	$-$3.8	&	2529	&	\ldots	&	Possible evol.	\\
J85.156861-70.153736	&       \ldots             &	05:40:37.6	&	$-$70:09:13	&	$-$3.8	&	2543	&	\ldots	&	Possible evol.	\\
J85.340519-69.529929	&       HN 5999            &	05:41:21.7	&	$-$69:31:48	&	$-$7.1	&	56957	&	\ldots	&	O-AGB	\\
J87.249164-70.556065	&       IRAS 05495--7034   &	05:48:59.7	&	$-$70:33:22	&	$-$5.3	&	10049	&	190	&	C-AGB	\\
  \hline
\multicolumn{7}{l}{Note: The HERITAGE-BMC designation prefix is HSOBMHERICC.}
 \end{tabular}
\end{table*}

\begin{table*} 
\centering
\caption{Details of the final sample of evolved star candidates in the SMC. For each source, the BMC name,  alias, BMC coordinates, bolometric magnitudes ($M_{\rm bol}$), total luminosity, SMC IRS identification (SMC IRS) and final classification is given.}
\label{tab:SMC_Fsample}
\centering
 \begin{tabular}{@{}lccccccc@{}}
   \hline
   \hline
BMC Source Name 	&  Alias    &	RA	&	Decl.	&	$M_{\rm bol}$	&	Luminosity 	&	SMC      	&	Classification 	\\
                  	&           & (J2000)	& (J2000)	&	          	& ($L_{\odot}$)		&	IRS         	&		\\
                                
\hline                          
J12.035323-73.248808	&   LIN 107            & 00:48:08.5	&	$-$73:14:56	&	$-$9.0	&	324932	&	\ldots	&	PN	\\
J12.150026-72.967348	&   LIN 115            & 00:48:36.0	&	$-$72:58:02	&	$-$6.2	&	24197	&	32	&	CPN	\\
J12.742224-73.86164	&   \ldots             & 00:50:58.1	&	$-$73:51:42	&	$-$4.6	&	5427	&	\ldots	&	Possible evol.	\\
J12.903066-73.338318	&   Jacoby SMC 17      & 00:51:36.7	&	$-$73:20:18	&	$-$3.3	&	1624	&	\ldots	&	B[e]	\\
J13.050807-73.147819	&   BMB-B 75           & 00:52:12.2	&	$-$73:08:52	&	$-$5.9	&	17561	&	177	&	O-AGB	\\
J13.579993-72.485754	&   \ldots             & 00:54:19.0	&	$-$72:29:09	&	$-$5.2	&	9946	&	260	&	Symbiotic star	\\
J20.652614-73.171815	&   \ldots             & 01:22:36.6	&	$-$73:10:19	&	$-$3.7	&	2404	&	\ldots	&	Possible evol.	\\
  \hline
\multicolumn{8}{l}{Note: The HERITAGE-BMC designation prefix is HSOBMHERICC.}
 \end{tabular}
\end{table*}

\begin{table*} 
\centering
\footnotesize
\caption{Summary of the columns present in the table of photometry which is available on-line only. \label{tab:photTabsummary}}
\begin{tabular}{lllc}
\hline
Column 	&	Column Name      &	Description	&	Null	\\
Number  &                        &                      &       Value    \\ 
\hline
1	&	SourceName	&	HERITAGE-BMC Designation	&	\ldots	\\
2	&	RA	&	Right Ascension (J2000)	&	\ldots	\\
3	&	Dec	&	Declination (J2000)	&	\ldots	\\
4       &       Galaxy  &       LMC or SMC              &	\ldots	\\
5-12	&	fU, dU, fB,dB, fV, dV, fI, dI  	&	MCPS U, B, V, I  fluxes (f) and errors (d) [mJy]	&	-999.9	\\
13-18	&	fJ, dJ, fH, dH, fK, dK 	&	2MASS J, H, K fluxes and errors  [mJy]	&	-999.9	\\
19-26	&	f3.6, d3.6, f4.5, d4.5, f5.8, d5.8, f8.0, d8.0 	& 	IRAC fluxes and errors [mJy]	&	-999.9	\\
27-30	&	f24, d24, f70, d70 	& MIPS fluxes and errors [mJy]	&	-999.9	\\
31-40	&	f100, d100, f160, d160, f250, d250,  f350, d350, f500, d500	& PACS \& SPIRE fluxes and errors [mJy]	&	-999.9	\\
41-45	&	flag100, flag160, flag250, flag350, flag500	& HERITAGE-BMC quality flags	&	-999.9	\\
46-53	&	fW1, dW1, W2, dW2, fW3,  dW3, fW4, dW4	&	WISE flux and errors [mJy]	&	-999.9	\\
54-63	&	fN3, dN3, fS7, dS7, fS11, dfS11, fL15, dfL15, fL22, dfL22 	&	AKARI flux and errors [mJy] 	&	-999.9	\\
64	&	SSID/ SMC IRS	& Spectral Identification number of the target from 	&	-999.9	\\
        &                       & \cite{Woods2011, Ruffle2015} and  &	\\
        &                       &  Woods et al. (in prep). &	\\
\hline
\end{tabular}
\end{table*}

\begin{table} 
\centering
\caption{Classification groups and tally for the LMC and SMC. \label{tab:classtally}}
\begin{tabular}{lc|lc}
\hline
\multicolumn{2}{c|}{LMC} & \multicolumn{2}{c}{SMC} \\
 Type & Count &  Type & Count \\
\hline
B[e]	&	3	&	B[e]	&	1	\\
BSG	&	1	&	O-AGB	&	1	\\
C-AGB	&	2	&	PN	&	2	\\
C-PAGB	&	1	&	Possible evol.	&	2	\\
LBV	&	1	&	Symbiotic star	&	1	\\
O-AGB	&	1	&		&		\\
O-AGB/PAGB	&	1	&		&		\\
PN	&	4	&		&		\\
Possible evol.	&	7	&		&		\\
RSG	&	3	&		&		\\
SNR	&	2	&		&		\\
UNK	&	1	&		&		\\
WR	&	1	&		&		\\
\hline
Total	&	28	&	Total	&	7	\\
\hline
\end{tabular}
\end{table}

\subsection{Comparisons to colour-magnitude diagrams}
\label{sec:CMDs}

\begin{figure} 
\centering
\includegraphics[width=3.5in]{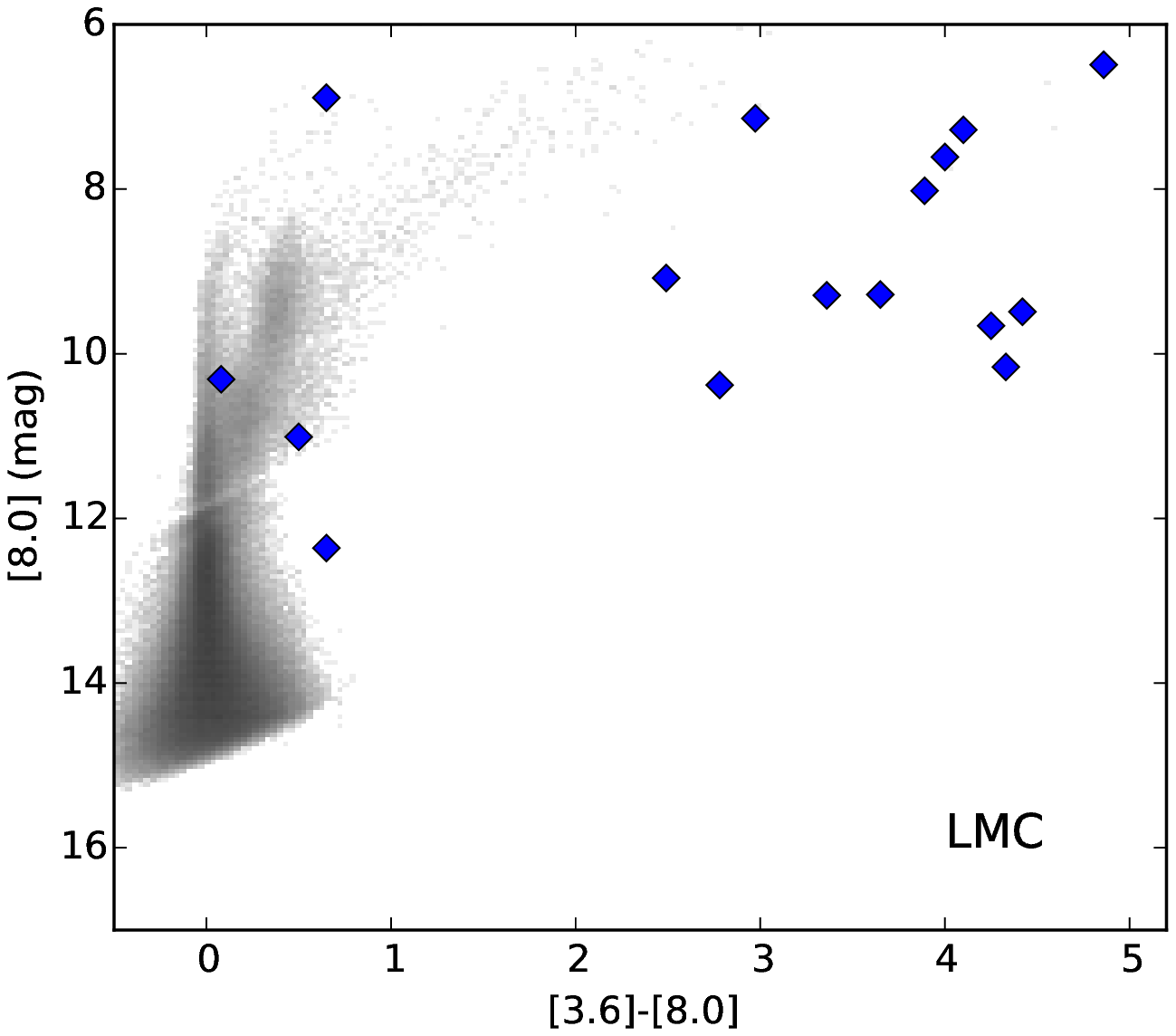}
\includegraphics[width=3.5in]{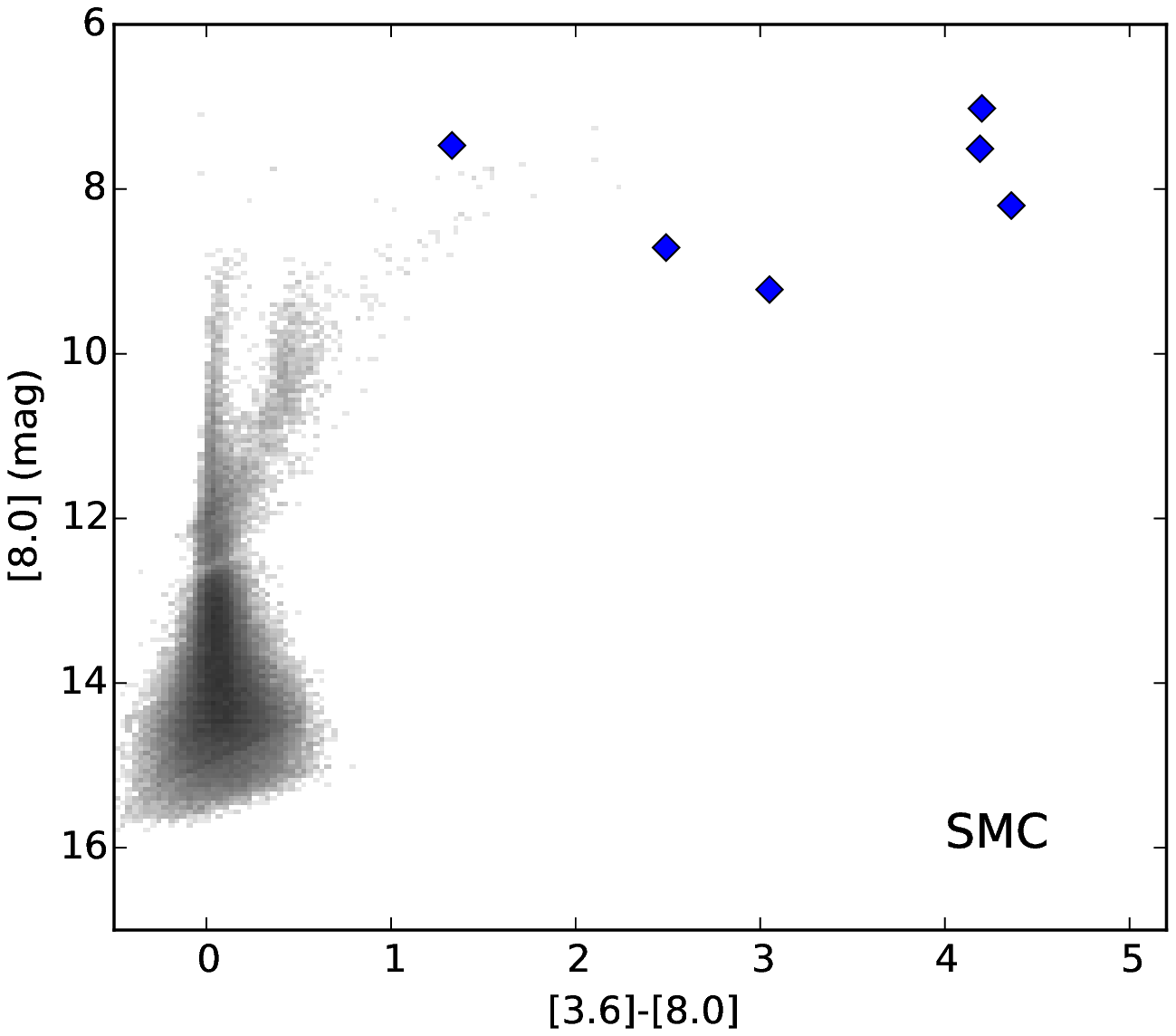}
 \caption[hessCMD]{The [8.0] vs. [3.6]--[8.0] CMD for the LMC (top) and SMC (bottom) showing the distribution of the post-main sequence stars in our final sample (blue diamonds) plotted over a Hess diagram of the evolved star sample of \cite{Boyer2011} for the LMC and SMC.} 
  \label{fig:hessCMDirac}
\end{figure}

\begin{figure} 
\centering
\includegraphics[width=3.5in]{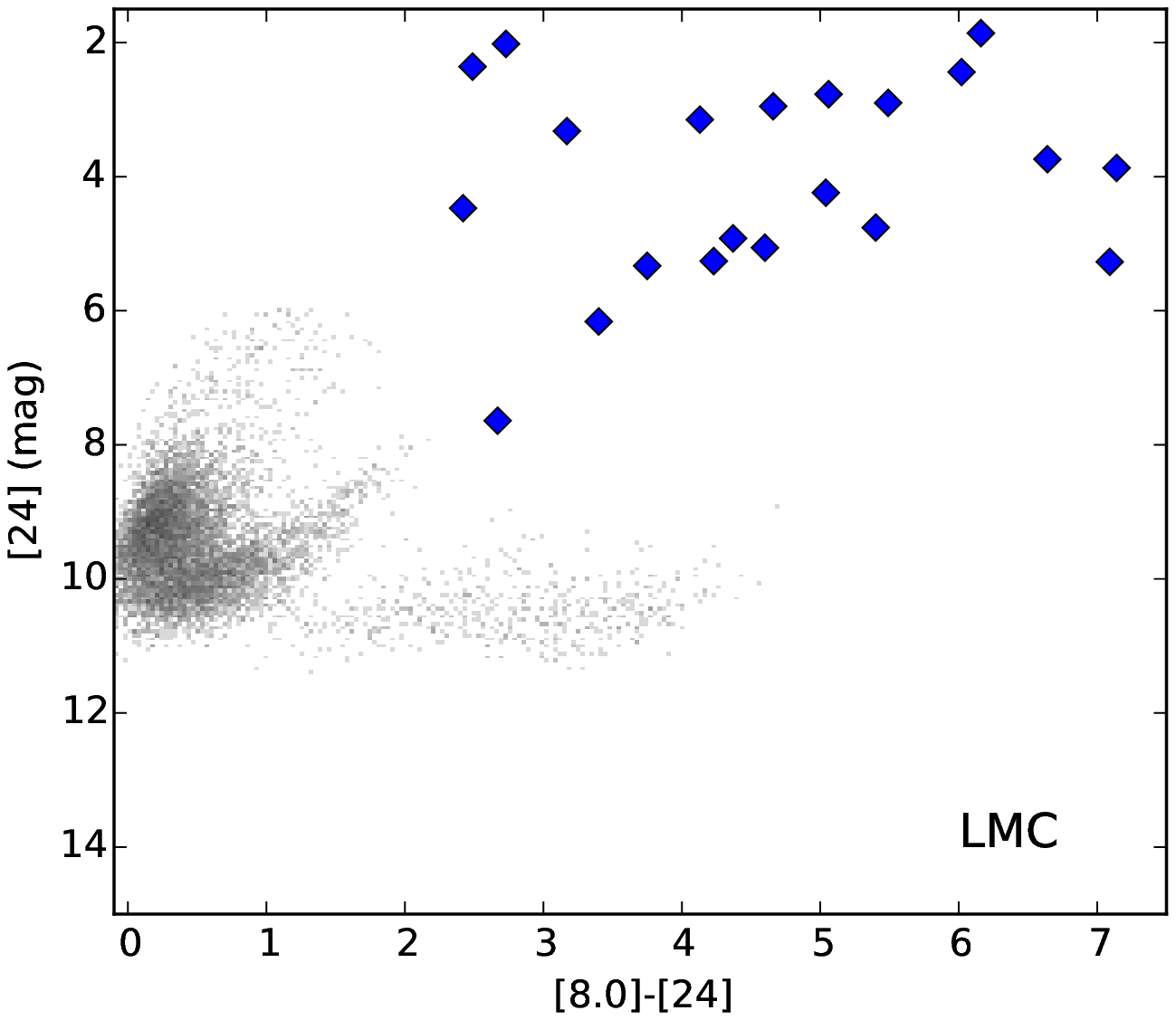}
\includegraphics[width=3.5in]{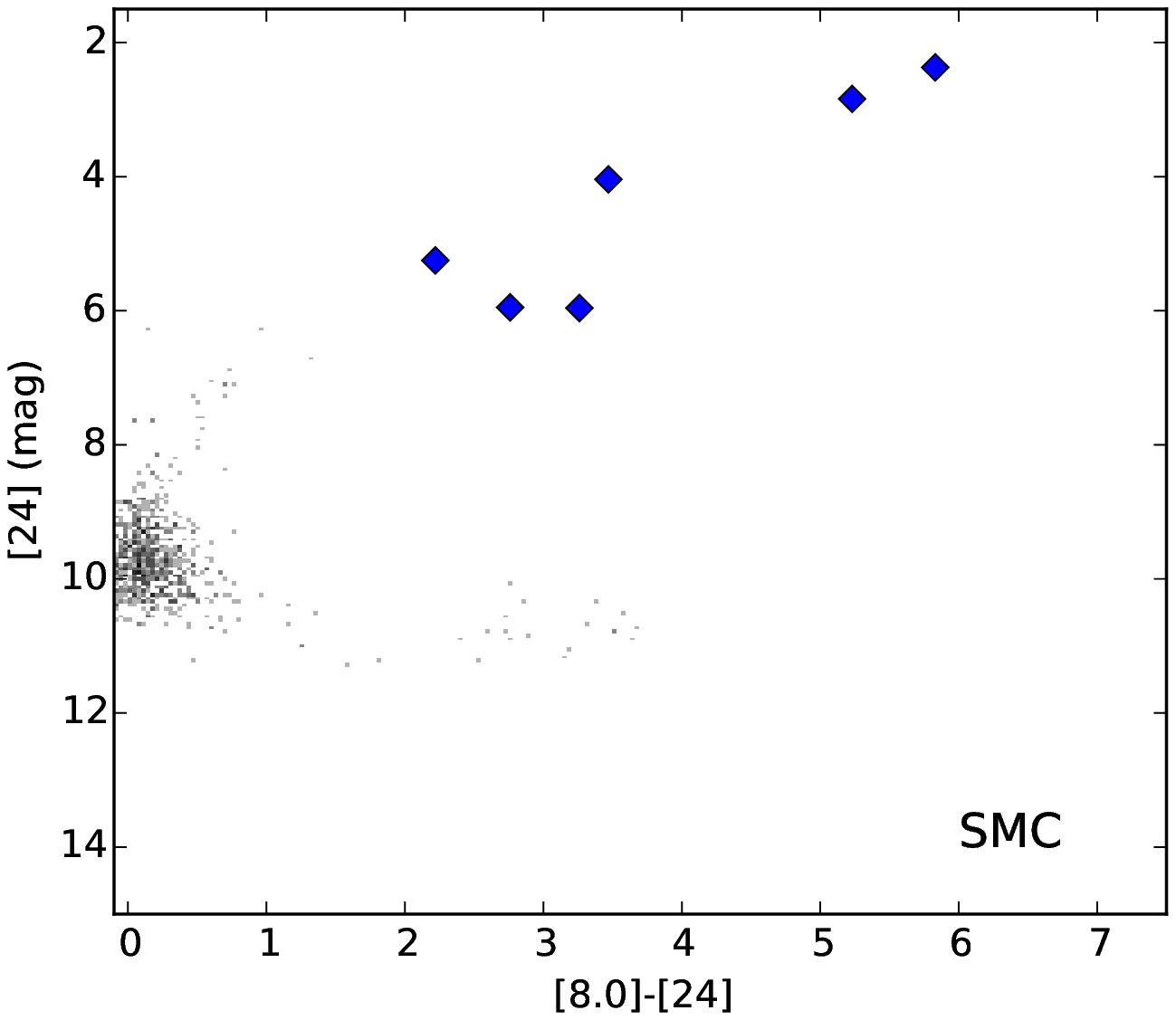}
 \caption[hessCMD]{The [24] vs. [8.0]--[24] CMD for the LMC (top) and SMC (bottom) showing the distribution of the post-main sequence stars in our final sample (blue diamonds) plotted over a Hess diagram of the evolved star sample of \cite{Boyer2011} for the LMC and SMC.} 
  \label{fig:hessCMDmips}
\end{figure}

IR colour-magnitude diagrams have been used to classify large numbers of objects detected by photometric surveys in the MCs \cite[e.g.][]{Blum2006, Matsuura2009, Boyer2011}. We compare the evolved sources with a {\em Herschel} detection to mid-IR colour-magnitude diagrams of the total evolved stellar population of the LMC and SMC. Figure~\ref{fig:hessCMDirac} shows the objects identified as candidate evolved stars detected with {\em Herschel} plotted as blue diamonds in the [8.0] versus [3.6]--[8.0] CMD; Figure~\ref{fig:hessCMDmips} shows them in the [24] versus [8.0]--[24] diagram. The stars in our sample tend to be offset/isolated from the majority of evolved star candidates identified from the SAGE sample; they are much redder in colour and have bright [8] and [24] magnitudes. Due to their very dusty nature they typically lie outside CMD cuts for evolved stars or reside in areas of high confusion.  

In the LMC there are four exceptions to this. These sources are identified as HSOBMHERICC J78.424908-67.403028 (BSDL 923), J78.608577-68.60194  (2MASS J05142617--6836085), J82.673312-69.193322  (HV 2605) and J85.340519-69.529929 (HV 5999) and are classified as either O-AGB/PAGB stars or blue supergiants. These stars are oxygen-rich and have SEDs that show a hot stellar component and a second peak at longer wavelengths due to a detached dust shell. Thus their [3.6]--[8.0] colour will resemble an evolved star with little circumstellar excess, while at [8.0]--[24] the contribution from circumstellar dust becomes significant and their colours resemble the other stars in our sample with a strong dust excess.

\subsection{Low and intermediate mass stars}

Our sample contains thirteen confirmed low to intermediate mass stars and PNe, which make up 32 per cent of the total sample. If we include the possible evolved stars (Section~\ref{sec:Unknown}), which have unknown mass, this value could rise to 60 per cent of the probable evolved sources detected in the HERITAGE-BMC. Due to their intrinsic brightness, stars with masses below $8 {\rm M}_{\odot}$ were not expected to be recovered by the HERITAGE survey, as only dust ejected by a few of the most massive stars were predicted to have luminosities greater than the {\em Herschel} sensitivity limits. 

The intermediate mass stars we detect include two of the most extreme AGB stars in the LMC with exceptionally high mass-loss rates, whose carbon rich chemistry has been confirmed spectroscopically \citep{Gruendl2008}. The SEDs of the two carbon stars (C-AGB) and the symbiotic (AGB-white dwarf binary) star do not show any evidence for excess emission in the far-IR, however. All the oxygen-rich AGB (O-AGB) stars in our sample show a stellar component, a component due to warm dust and in some instances a far-IR excess may also be present. This excess may be associated with the dust-driven wind, a stable circumstellar disc or dust swept-up from the surrounding ISM. The mass-loss rate stated for the O-AGB stars is likely an underestimate as current model grids do not account for complex dust geometries due to a detached shell. Detailed 3D radiative-transfer modeling of the sources with a far-IR excess is required to determine reliable dust production rates and to differentiate between ISM dust and dust produced in the stellar wind (Jones et al., in preparation, Paper II).  

Our final sample also includes several highly evolved post-superwind objects such as planetary nebulae. The five planetary nebulae in our sample have been identified from a combination of H$\alpha$ images, optical ionized gas emission lines and mid-IR spectra. Due to the similarities between the SEDs of PNe and YSOs and the strict criteria we applied when selecting the evolved star sample it is likely that we have excluded some PNe from our final list. Thus the number of PNe with a {\em Herschel} counterpart included in our final catalog should be treated as a lower limit.

\subsubsection{AGB stars}
\label{sec:AGB}

\noindent {\em HSOBMHERICC J78.608577-68.60194 (2MASS J05142617--6836085):} This long period variable (P = 382 d; \citealt{Fraser2008}) is likely oxygen-rich. It was found to have a mass-loss rate (MLR) of  $2.8\times10^{-7}$ ${\rm M}_{\odot} \, {\rm yr}^{-1}$ \citep{Riebel2012}. The SED is double peaked which may indicate that the source is a post-AGB star or that it has an unresolved detached shell. 

\noindent {\em HSOBMHERICC J82.683128-71.716888 (IRAS 05315--7145)} is an extreme carbon-rich AGB star, heavily obscured by dust. It has a very high mass-loss rate of $1.7\times10^{-4}$ ${\rm M}_{\odot} \, {\rm yr}^{-1}$ \citep{Gruendl2008}; this is the second highest of all the extreme AGB (x-AGB) stars in the LMC. SiC absorption features are present in the IRS spectra. 

\noindent {\em HSOBMHERICC J85.340519-69.529929 (HV 5999):}  This source has a luminosity of  $M_{\rm bol} = -7.1$ mag and a spectral type of M3 suggesting the source is likely a massive O-rich AGB star undergoing hot bottom burning or is a red supergiant. It is also included in the RSG catalog of \cite{Massey2003}, who give $M_{\rm bol} = -8.05$ mag.  Fits with GRAMS indicate that this source is oxygen-rich with a mass-loss rate of $1.4\times10^{-6}$ ${\rm M}_{\odot} \, {\rm yr}^{-1}$ \citep{Riebel2012}. The SED shows evidence of a far-IR excess which (if associated with the star) may indicate the presence of large (\mum -- mm size) grains and/or cold dust in a detached shell. 

\noindent {\em HSOBMHERICC J87.249164-70.556065 (IRAS 05495--7034)} is an extremely red carbon star first identified by \cite{Gruendl2008}. Its spectra show the presence of amorphous carbon, SiC absorption features, and a strong MgS feature at 30 $\mu$m. This `extremely red object' (ERO) has the highest mass-loss rate ($2.3\times10^{-4}$ ${\rm M}_{\odot} \, {\rm yr}^{-1}$) of all the x-AGB stars in the LMC \citep{Gruendl2008}.


\noindent {\em HSOBMHERICC J13.050807-73.147819 (BMB-B 75)} is a Mira-type variable with a long period of P = 1453 d \citep{Cioni2003}. It is an M6-type star \citep{Blanco1980}. The star is oxygen-rich with a mass-loss rate of $1.5\times10^{-5}$ ${\rm M}_{\odot} \, {\rm yr}^{-1}$ \citep{Jones2012}. Recently a redshifted galaxy (z$\sim$0.16) has been discovered in the same line of sight as BMB-B 75 (Kraemer et al. 2015, in prep.) The far-IR emission detected by {\em Herschel} is probably due to this spatially coincident galaxy.

\noindent {\em HSOBMHERICC J13.579993-72.485754	(SSTISAGEMA J005419.21-722909.7)} is a D-type symbiotic star, consisting of an AGB star and white dwarf \citep{Oliveira2013}. Of the seven known symbiotic stars in the SMC it is the only one with warm dust; this dust is silicate rich and unprocessed as it shows no PAH emission.

\subsubsection{Post-AGB stars}
\label{sec:PAGB}

Our sample only includes one confirmed post-AGB star. However \cite{Seale2014} notes that the BMC contains four confirmed post-AGB stars from \cite{vanAarle2011} with low resolution optical spectra. Of these four sources, both HSOBMHERICC J73.375205-69.296953 and J78.854807-66.317713 are included in our sample as a WR star and a probable evolved star, respectively (see Sections~\ref{sec:Massivestars} and \ref{sec:Unknown}). However, HSOBMHERICC J80.919272-68.090965 and J83.220747-69.987453 were eliminated from our sample either because the SED resembles that of a YSO or because the SED at longer wavelengths appears disjointed and the {\em Herschel} data were attributed to a mismatch. The SEDs of these two sources are shown in Figure~\ref{fig:egSEDs_PAGB}.

\noindent {\em HSOBMHERICC J79.617143-68.067639 (IRAS 05185--6806)} was first identified as a likely post-AGB object by \cite{vanLoon1997}. Its mid-IR {\em Spitzer}-IRS spectrum shows the presence of the unidentified 21-\mum feature found only in carbon-rich post-AGB stars \citep{Volk2011, Matsuura2014, Sloan2014}, confirming that mass loss during the AGB phase has terminated.

\subsubsection{PNe}
\label{sec:PNe}

\noindent {\em HSOBMHERICC J76.216758-68.652343 (SMP LMC 21)} is a very high-excitation, quadrupolar PN of Type I \citep{Stanghellini1999}. It has an extremely strong IR-excess (T$_{\rm dust \, cont} = 130$ K) and the {\em Spitzer} IR spectrum shows prominent crystalline silicate emission due to forsterite and enstatite. No other solid-state features are seen \citep{Stanghellini2007}. The spectrum also shows both low and very high excitation nebular emission lines (i.e., [S {\sc iii}] and [Ne {\sc vi}], respectively). It is one of the reddest PNe in the Magellanic Clouds. The ionized mass in the nebula is determined to be between 0.2 -- 0.6 ${\rm M}_{\odot}$ \citep{Barlow1987, Boffi1994}, and the progenitor is thought to be a relatively massive AGB star.

\noindent {\em HSOBMHERICC J76.989649-68.863313 (SMP LMC 28)} is a high-excitation, isolated, oxygen-rich PN \citep{Reid2006, Hora2008, BernardSalas2009}. It is has a high nitrogen and neon abundance compared to other PNe \citep{Stanghellini2005, BernardSalas2008}. 
This PN is resolved in HST images and has three co-aligned round shapes, corresponding to the central star and two faint arms \citep{Shaw2001}. The central star was detected by \cite{Villaver2003}. They estimate its luminosity to be $L = 1.4 \times 10^{4} {\rm L}_{\odot}$, with ${\rm T}_{\rm eff} \lesssim 17000$ K, which corresponds to a progenitor mass of $\sim 3{\rm M}_{\odot}$. The ionized mass in the nebula is $<0.2 {\rm M}_{\odot}$ \citep{Wood1987}.

\noindent {\em HSOBMHERICC J77.014627-68.671229  (LHA 120-N 102/ SMP LMC 29)} is a bright PN noted to have strong emission lines \citep{Reid2006}.  This PN is resolved and has an irregular shape with a bipolar core in HST images. It is thought to be one of the smallest bipolar PNe recognised to date, with radius of $\sim$0.27 arcsec in [O {\sc iii}] \citep{Shaw2006}. No central star is evident within the bright nebulosity of the image \citep{Villaver2003}. 
It has emission lines in high states of ionization (e.g., N {\sc v}) which have extreme photoionization temperatures (T $\sim$ 200 kK) \citep{Herald2007}.


\noindent {\em HSOBMHERICC J12.150026-72.967348 (Lin 115/ SMP SMC 11)} is a well known carbon-rich PN which was first discovered by \cite{Henize1956}. Both emission line observations and PAH features in its IRS spectrum confirm its carbon-rich nature \citep{Stanghellini2003, BernardSalas2009}. High-resolution HST images show that it has a complex bipolar morphology, which extends out to 2$''$ from the center \citep{Stanghellini2003}. It has strong mid-IR emission; the dust continuum sharply rises after 12 \mum and its SED peaks around 35 $\mu$m, indicative of a far-IR-excess \citep{BernardSalas2009}. 

\noindent {\em HSOBMHERICC J12.035323-73.248808 (LIN 107)} was identified as a planetary nebula by \cite{Lindsay1961, Meyssonnier1993}. Other authors list this object as a weak emission-line object or `low-excitation blob' \citep{Meynadier2007, Sheets2013, Testor2014} suggesting that it is an embedded star cluster or compact \Hii region.

\subsection{Massive stars}
\label{sec:Massivestars}

We have compared the HERITAGE-BMC to the {\em Spitzer} catalogs of massive stars in the Magellanic Clouds \citep{Bonanos2009, Bonanos2010} and find nine matches in the LMC and one match in the SMC. This is approximately twice the number anticipated prior to the HERITAGE observations.  
Notably far-IR emission is observed for four of the fifteen confirmed B[e] stars in the Magellanic Clouds, while only one of the  106 late-type WR stars has a far-IR detection. 

Three heavily dust enshrouded extreme RSGs are also detected. These stars all optically thick in the mid-IR where amorphous silicate absorption features are seen at 10 and (in some instances) $\rm {20{\ }\mu}$m indicative of extremely high mass-loss rates, which may be as large as ${10}^{-3}{\ } \rm{M}_{\odot}~{\rm yr}^{-1}$ \citep{Dijkstra2003}. The mass loss from these three sources was missed in dust budget estimates for the LMC by \cite{Riebel2012}.

\subsubsection{BSG stars}

\noindent {\em HSOBMHERICC J78.424908-67.403028 (BSDL 923)} is a blue supergiant (BSG); a massive star, newly evolved from the main sequence. Its location in the young cluster LMC-N30 \citep{Gouliermis2003, Bica1999} is consistent with that of a massive star. The SED is distinctly double-peaked, falling to a minimum at $\sim$8 $\mu$m before rising steeply long-ward of 15 $\mu$m; this red peak is due to a circumstellar disc or torus. Weak silicate features are seen in its {\em Spitzer}-IRS spectrum \citep{Woods2011}.

\subsubsection{RSG stars}

\noindent  {\em HSOBMHERICC  J73.793532-68.341596 (WOH G064)} is a very luminous red supergiant with spectral type M7.5 \citep{Elias1986}  and an effective temperature of ${\rm T}_{\rm eff} \sim3200$ K. It is surrounded by a thick dust shell and is one of six evolved stars in the Magellanic Clouds to have a 10 \mum amorphous silicate feature in self-absorption \citep[][Jones et al.~(in prep)]{Roche1993, Trams1999, Buchanan2006, Kemper2010}. Its mass-loss rate is estimated to be in the region of $10^{-3}$ ${\rm M}_{\odot} \, {\rm yr}^{-1}$ \citep{vanLoon2005} with a wind speed of $v_{\rm wind} \sim25 {\rm km s}^{-1}$, determined from OH, SiO, and H$_2$O maser emission \citep{Wood1986, vanLoon1996, vanLoon1998, Marshall2004}. The circumstellar dust shell was resolved by \cite{Ohnaka2008}, who found that the envelope had a bipolar geometry; they derived a circumstellar envelope mass of 3--9 ${\rm M}_{\odot}$. 

We estimate the bolometric luminosity to be $4\times10^{5}$ ${\rm L}_{\odot}$ ($M_{\rm bol}$ $=$ $-$9.3), however if the geometry of the dust shell is non spherical then the bolometric luminosity is an overestimate. Accounting for the non spherical nature of the shell \cite{Ohnaka2008} determine the luminosity to be $2.8\times10^{5}$ ${\rm L}_{\odot}$ ($M_{\rm bol}$ $=$ $-$8.8)  leading to a down-ward revision of the star's initial mass to ${\rm M}_{\rm ZAMS} \sim25 {\rm M}_{\odot}$. Its location beyond the Hayashi limit in the H-R diagram indicates that the star is no longer in hydrostatic equilibrium \citep{Levesque2007} and it may be experiencing an intense unstable phase of heavy mass loss. \cite{vanLoon2010} suggests that the mass-loss rate of WOH G064 has increased dramatically in the past few thousand years (i.e. it has experienced the onset of the superwind phase), as they find no evidence for a submm excess. 

\noindent  {\em HSOBMHERICC J81.920736-69.135316 (IRAS 05280--6910)} is a highly obscured OH/IR red supergiant with an initial mass ${\rm M}_{\rm ZAMS} \sim 20{\rm M}_{\odot}$ \citep{Wood1992}.  Its stellar counterpart was identified using high-resolution near- and mid-IR imagery by \cite{vanLoon2005b}. OH maser emission has been used to determine its wind outflow velocity which is  $\approx20 {\rm km s}^{-1}$ \citep{Marshall2004}; the star also exhibits H$_2$O maser emission \citep{vanLoon2001}. 
It has an extremely high mass-loss rate of $\dot{M}\simeq2-8 \times 10^{-3}$ ${\rm M}_\odot \, {\rm yr}^{-1}$ \citep{Boyer2010} and is potentially the most dust-enshrouded supergiant in the Magellanic Clouds. In the IRS spectra both the 10 and 20 \mum amorphous silicate features are seen in absorption  and a declining, featureless continuum is seen at far-IR wavelengths \citep{Kemper2010}.  The dust shell may have a flattened geometry which (unlike WOH G064) is viewed edge-on. The total dust mass is $\approx$ $0.3~{\rm M}_\odot$. 
There is no evidence of excess emission at $\lambda > 100$ $\mu $m, implying that there is no significant contribution from dust colder than $\sim$100 K or from large ($\mu $m - mm size) grains in a detached shell \citep{vanLoon2010, Boyer2010}. 

\noindent  {\em HSOBMHERICC J83.558492-69.789067 (IRAS 05346-6949):} This luminous oxygen-rich red supergiant is highly obscured \citep{Elias1986}; due to its very red IRAC [3.6]--[4.5] $\sim$ 2.5 colour it has been  classified as an ERO by \cite{Gruendl2009} and an extreme-AGB star by \cite{Srinivasan2009}. We classify it as a RSG because of its high luminosity and the silicate features present in its IRS spectrum \citep{Jones2014}. Unlike many sources in the ERO class this source is not carbon-rich, instead it has an oxygen-rich dust composition. 
Like WOH G064, the {\em Spitzer}-IRS spectrum of {\em HSOBMHERICC J83.558492-69.789067} exhibits a 10 $\mu $m silicate feature in absorption \citep[][Woods et al. (in prep), Jones et al. (in prep)]{Jones2012} which means that the dust envelope is optically thick, and the star is experiencing heavy mass loss.  Despite being bright in the mid-IR, any maser emission has not been detected \citep{vanLoon2001} and the source is not bright in the optical. Radio continuum emission at 1420 MHz may be associated with the source.

\subsubsection{B[e] stars}

\noindent  {\em HSOBMHERICC J74.195411-69.840206 (R66)} is a well-known B[e] supergiant of spectral type B8 Ia, with luminosity $L = 3 \times 10^{5}$, ${\rm T}_{\rm eff} \sim 12000$ K and an initial mass of ${\rm M}_{\rm ZAMS} \sim 30{\rm M}_{\odot}$  \citep{Lamers1998b, Stahl1983}. The star is both cool and below the Humphreys-Davidson limit \citep{Zickgraf1986}, leading \cite{vanLoon2010} to suggest that it is a blue-loop star on its way back to becoming an RSG. The {\em Spitzer}-IRS spectrum shows PAHs, amorphous silicates and strong crystalline silicate features, indicative of dust grain processing in a long-lived pole-on circumstellar disc \citep{Kastner2006, Kastner2010, Aret2012}. This disc is massive and contains $\approx$ $0.001~M_\odot$ of dust.

\noindent  {\em HSOBMHERICC J82.673312-69.193322 ([BE74] 328)} is a luminous ($M_{\rm bol}$ $=$ $-$10.15; \citealt{Massey2002}), emission line star of spectral type B1.5IIe \citep{Reid2012}. It is listed as a poorly studied eruptive variable of early spectral type in the General catalog of Variable Stars \citep{Samus2004}. 

\noindent  {\em HSOBMHERICC J84.10768-69.381998  (R126)} is an extremely luminous B[e] supergiant in the LMC. Its bolometric luminosity is $L = 1.2 \times 10^{6} {\rm L}_{\odot}$, with ${\rm T}_{\rm eff} \sim 22500$ K and is estimated to have an initial mass of ${\rm M}_{\rm ZAMS} \sim 75{\rm M}_{\odot}$. Its present day mass is ${\rm M} \sim 40{\rm M}_{\odot}$ \citep{Zickgraf1985}. 
Estimates for the current mass-loss rate range between $\dot{M} = 10^{-6} - 4.6 \times 10^{-5}$ $M_{\odot}$ ~yr$^{-1}$, depending on the assumed terminal velocity of the stellar wind $v_\infty \approx 650 - 1800 $~km~s$^{-1}$ \citep{Zickgraf1985, Bjorkman1998}. 
R 126 is thought to have a relatively massive `puffed up' circumstellar disc which is estimated to contain ${\rm M}_{\rm dust} \approx$ $0.003~{\rm M}_\odot$ of dust \citep{Kastner2006}. Detailed analysis of R 126's disc can be found in \cite{Porter2003, Kraus2007}.


\noindent {\em HSOBMHERICC J12.903066-73.338318 (Jacoby SMC 17)} is included in the list of probable B[e] supergiants in the SMC \citep{Mennickent2002}, as its light curve resembles Galactic B[e] stars. \cite{Meyssonnier1993} identify the source as a very low excitation object with a blue continuum. It was also identified as an emission line star by \cite{Murphy2000}. It may potentially be a PN \citep{Boroson1989}; its total luminosity is on the low-side for a B[e] star but is consistent with a PNe.

\subsubsection{LBV stars}

\noindent {\em HSOBMHERICC J75.5321-71.337365 (R71)} is one of six confirmed dusty luminous blue variables (LBVs) in the LMC \citep{Humphreys1994}. Its initial mass is estimated to be ${\rm M}_{\rm ZAMS} \sim40 {\rm M}_{\odot}$  \citep{Lennon1993}.
LBVs can develop instabilities and show sudden bursts of mass loss; this mass-loss history can be probed via dusty ring nebulae often associated with the star \citep{Hutsemekers1994}.  R71 recently experienced an eruption, where there was a $\sim$2 mag increase in the V-band brightness \citep{Szczygiel2010}. This increase in emission is not seen in our mid-IR data which show a smooth transition between the {\em Spitzer} and {\em Herschel} photometry, even-though the {\em Herschel} data were obtained post-eruption, while the {\em Spitzer} observations were obtained prior to the outburst. The difference in flux between the {\em WISE} bands at 3.4 and 4.6 \mum and the IRAC bands may be due to the increased emission from the photosphere during the recent outburst. 

The dust shell of R 71 contains PAHs, and both amorphous and crystalline silicates indicative of non-equilibrium chemistry. It estimated to contain 0.02 ${\rm M}_{\odot}$ of dust, ejected at a rate of $7 \times 10^{-6}$ ${\rm M}_{\odot} \, {\rm yr}^{-1}$ over the past 3000 years. A second dust shell from an earlier expulsion event is estimated to contain 0.3 ${\rm M}_{\odot}$ of dust \citep{Voors1999}. This second dust shell has been called in to question by \cite{vanLoon2010} and \cite{GuhaNiyogi2014} who find that R 71 lacks dust colder than $T_{\rm d} \lesssim 100~{\rm K}$. 

Using data from the HERITAGE Science Demonstration Program \cite{Boyer2010} measured a far-IR dust component at $\lambda>250$ $\mu $m. In later HERITAGE photometric data-sets, which better accounts for background emission, the SPIRE 350 and 500 \mum is confirmed to be diffuse. Using this refined photometry, \cite{GuhaNiyogi2014} have fit radiative transfer models to R71 and derive a total dust mass of 0.01 ${\rm M}_{\odot}$ and a time-averaged dust mass-loss rate of $2.5\times10^{-6}$ ${\rm M}_{\odot} \, {\rm yr}^{-1}$.

\subsubsection{WR stars}

{\em HSOBMHERICC J73.375205-69.296953 (Brey 3a/ BAT99-4 )} is a WC9-type Wolf-Rayet (WR) star with a progenitor mass believed to be in excess of 25 ${\rm M}_{\odot}$ \citep{Breysacher1999}.  Late-type WR stars can lose mass at rates of up to  $10^{-4}$ ${\rm M}_{\odot} \, {\rm yr}^{-1}$ \citep{Crowther2007}. 
Characterised by strong He, N and C emission lines and red SEDs due to warm circumstellar dust, WR stars are relatively easy to identify; there are thought to be 134 WR stars in the LMC.  Brey 3a is unusual as its exact spectral type remains controversial, leading some authors to dispute its WR status and instead suggest that it may be a transition object between an Of star and a WR star \citep{Moffat1991, Heydari-Malayeri1992}.
\cite{Egan2001} note the presence of an M3-type red giant 5.2$''$ away from Brey 3a, which may contribute some flux at $<$24 $\mu$m but \cite{vanLoon2010} find that the far-IR emission seen in the MIPS-SED spectra of Brey 3a is typical of a WR-type star rather than an M-type giant with warm circumstellar dust. The IRS spectrum, which rises steeply at $\sim$15 $\mu$m, also resembles that of the WR-type star \citep[]{Woods2011}.

\subsection{Supernova remnants}
\label{sec:SNe}

Our {\em Herschel} sample includes two well known supernova remnants: 

\noindent A detailed analysis of SN 1987A {\em (HSOBMHERICC J83.866283-69.269995)} in the far-IR/sub-mm was conducted by \cite{Matsuura2011, Matsuura2015, Indebetouw2014}, who found that the SN ejecta contained a large reservoir of freshly-synthesised cold dust grains with a temperature of $\sim$20 K and a dust mass of 0.4--0.8 ${\rm M}_{\odot}$. This dust mass (measured 24 years after the explosion) is significantly higher than the $10^{-4} {\rm M}_{\odot}$ of dust reported in the first couple of years following the explosion \citep{Wooden1993}; providing the first clear evidence that supernovae can produce significant amounts of dust and that they can be an important source of dust in the ISM of galaxies. The dust produced in the ejecta is thought to be composed of silicates and amorphous carbon. 
As the core-collapse SN is rapidly evolving and the expanding ejecta have yet to interact with the warm ($\sim$170 K) circumstellar dust ring (ejected from its red supergiant progenitor; \citealt{Barkat1988}) and experience reverse shocks which may potentially destroy or alter the grains \citep[e.g.,][]{Jones1994, Lakicevic2015, Temim2015}, the total quantity of dust ejected into the ISM by SNe remains uncertain.

\noindent {\em HSOBMHERICC J81.264647-69.642784 (N 132D)} is the remnant of a core-collapse supernova that exploded $\sim$2700 years ago \citep{Williams2006}. The supernova remnant was detected as a slightly extended point source in two {\em Herschel} bands. Unlike SN 1987A, the emission from N 132D at these wavelengths is too faint to accurately constrain the dust mass of the ejecta.
N132D is one of the brightest X-ray remnants in the Magellanic Clouds, and its extended emission has been extensively studied in optical and IR wavelengths \citep{Morse1995}; the bright mid-IR emission has been attributed to swept-up ISM dust collisionally heated by the hot X-ray emitting plasma generated by SN blast waves \citep{Williams2006}. {\em Spitzer} observations also reveal large polycyclic aromatic hydrocarbons (PAHs) that have been processed by thermal sputtering in the blast wave \citep{Tappe2006}.   
Dust created in the SN ejecta has been observed by \cite{Rho2009}, via IR spectral lines; they estimate the mass of the newly formed dust in the ejecta to be $>$0.008 ${\rm M}_{\odot}$. 

The far-IR emission from other Magellanic Cloud supernova remnants \citep[see][and references therein]{Otsuka2010, Lakicevic2015} are not included in the HERITAGE point source catalogs as the dust emission from these remnants is either too faint or it is extended and large enough to be resolved by the {\em Herschel} observations.

\subsection{Possible evolved stars}
\label{sec:Unknown}

Two sources in our sample have low signal-to-noise or very unusual/unidentifiable features in the corresponding {\em Spitzer}-IRS spectra. Due to this uncertainty \cite{Woods2011} classified these sources as unknown objects. As the true nature of these objects has not been spectroscopically confirmed, we assign these objects a possible evolved star status. 

\noindent {\em HSOBMHERICC J74.728949-68.844596} is probably a planetary nebula (RP 1805); it has a diameter in H$\alpha$ of 5 arcsec \citep{Reid2006}. The mid-IR IRS-spectrum of the source shows no clear identifying features above the noise and was thus reported as a non-detection. The source is faint and appears elongated the SAGE-LMC 8-\mum images. A UV-bright star located within 1 arcmin of the source could conceivably be irradiating the dust cloud.

\noindent  {\em HSOBMHERICC J79.075912-71.899396  (IRAS 05170-7156)} is an unusual object. It was assigned an unknown (UKN) spectral type by \cite{Woods2011}. The mid-IR spectrum rises toward longer wavelengths and is featureless apart from a dip in the continuum between 11.5--16.0 $\mu$m. Its SED is indicative of a post-main sequence object, however, it has also been considered to be an AGN candidate \citep{deGrijp1987, Kozlowski2009} and a high-probability Stage 1 YSO \citep{Whitney2008}.

Nine sources in our sample are relatively unstudied in the literature, but detailed examination of the SEDs and fits with YSO models suggest that these sources are likely evolved stars. Only one source HSOBMHERICC J84.173847-66.808723 was given a less certain classification due to its unusual multi-peaked SED and other characteristics which meant that an alternate classification could not conclusively be ruled out.

\subsection{Physical extent of dust shells}
\label{sec:Detached shells}

Bow shocks due to wind-ISM interactions and detached shells around AGB stars are commonly seen in the Milky Way via their far-IR excess emission \citep{Kerschbaum2010, Groenewegen2011, Cox2012}. High-mass evolved stars are often associated with a ring nebula formed from either past stellar eruptions or by swept up ISM material. In all instances the maximum physical size observed between the central star and the detached shell or bow shock is smaller than 1 pc \citep{Cox2012}. These size scales are much smaller than the highest achievable resolution of 8.6--8.8$''$, 2--3 pc at 100 $\mu$m in the HERITAGE images. 

Given the physical size of the {\em Herschel} beam, we cannot use the images to distinguish between dust formed in the ejecta or dust swept-up from the surrounding ISM. Instead, the spectral energy distributions may provide some indication of far-infrared excess emission. The circumstellar envelopes of the extreme carbon stars, RSGs, LBV and WR stars in our sample lack a far-IR excess, and thus are unlikely to have large (\mum -- mm size) grains or harbor any significant reservoirs of cold dust. However, we may see excess emission (i.e. above what is expected from a circumstellar shell) from some of the oxygen-rich evolved stars, planetary nebulae and B[e] supergiants. This emission may either be due to dust produced by the evolved star or it may arise from swept-up ISM material. 

Dust mass estimates obtained using graybody fits to the far-IR emission can in principle be used to identify the likely source of this emission \citep[e.g.][]{Seale2014}, however this method is unlikely to yield reliable dust masses due to the complex geometry of the objects in our sample. Instead 3D radiative transfer modeling is required to derive reliable dust masses and temperatures (Jones et al. in prep, Paper II).
High-resolution follow-up observations with the Atacama Large Millimeter/submillimeter Array (ALMA) should also uncover the true source of this far-IR emission.  Followup observations will probe the mass-loss rate history of the star or the wind-ISM interaction at low metallicity.

\subsection{Selection effects and completeness}
\label{sec:completeness}

\begin{figure} 
\centering
\includegraphics[width=0.5\textwidth]{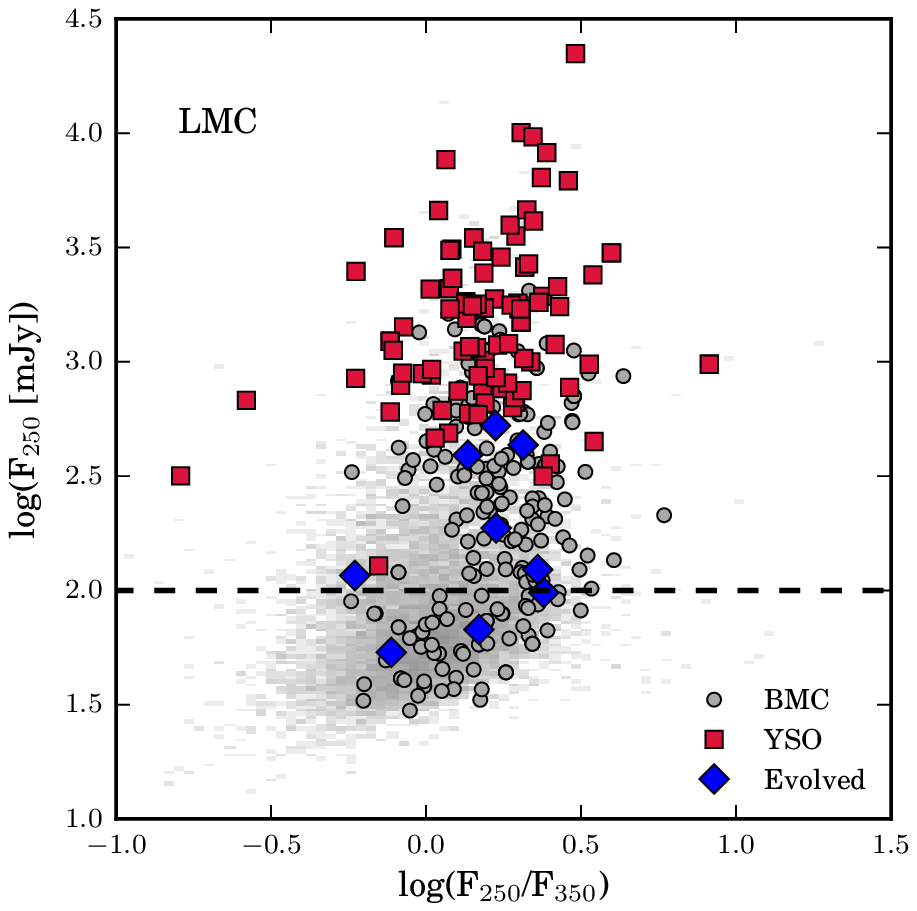}
\includegraphics[width=0.5\textwidth]{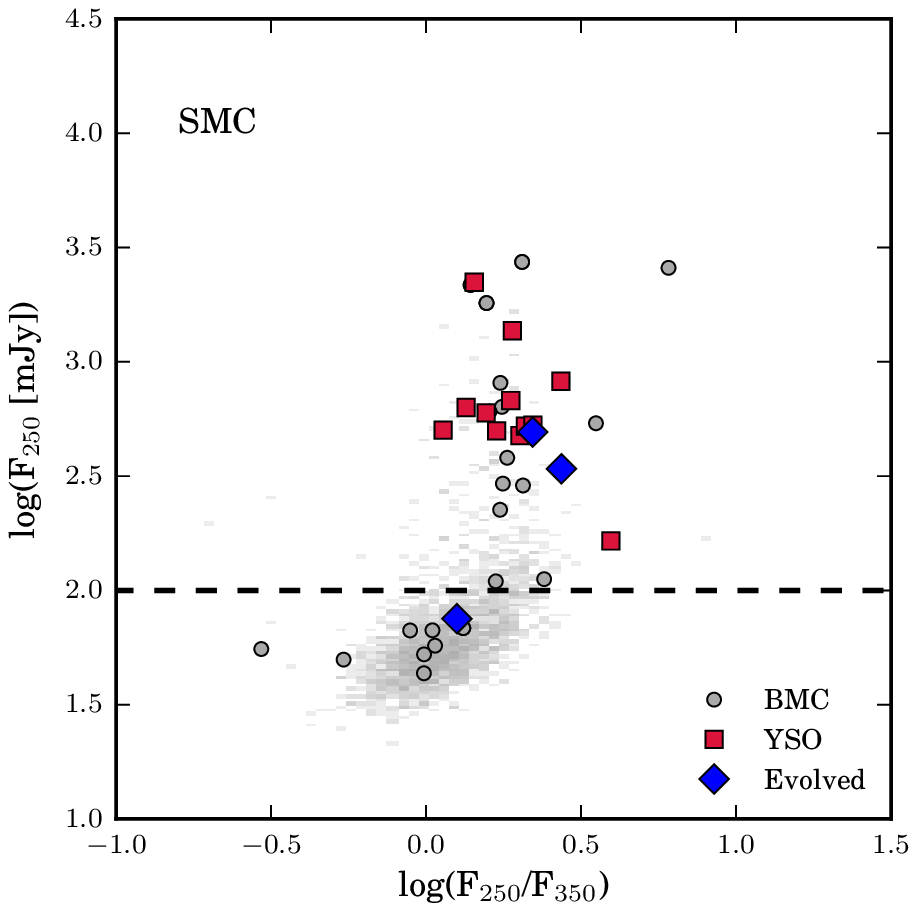}
 \caption[FIR CMDS]{SPIRE $F_{250}$ and $F_{350}$ CMD for the confirmed evolved sources (blue) in the LMC (top) and SMC (bottom) overlaid on a Hess diagram of the HERITAGE Band-Matched Catalog sources. The grey circles represent the initial list of BMC candidate evolved stars and the red squares show the candidate evolved stars which have been confirmed spectroscopically as YSOs. The dashed line marks the completeness limit of the BMC at low backgrounds.}
  \label{fig:FIRCMD}
\end{figure}

\begin{figure} 
\centering
\includegraphics[width=0.5\textwidth]{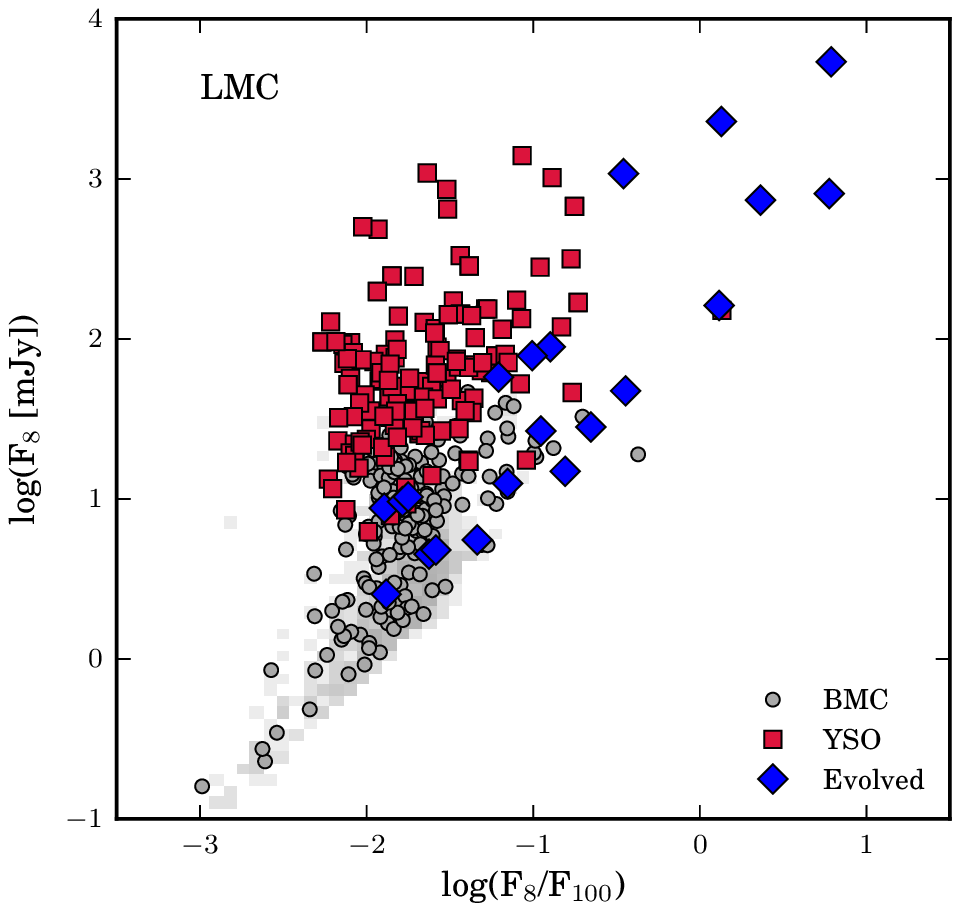}
\includegraphics[width=0.5\textwidth]{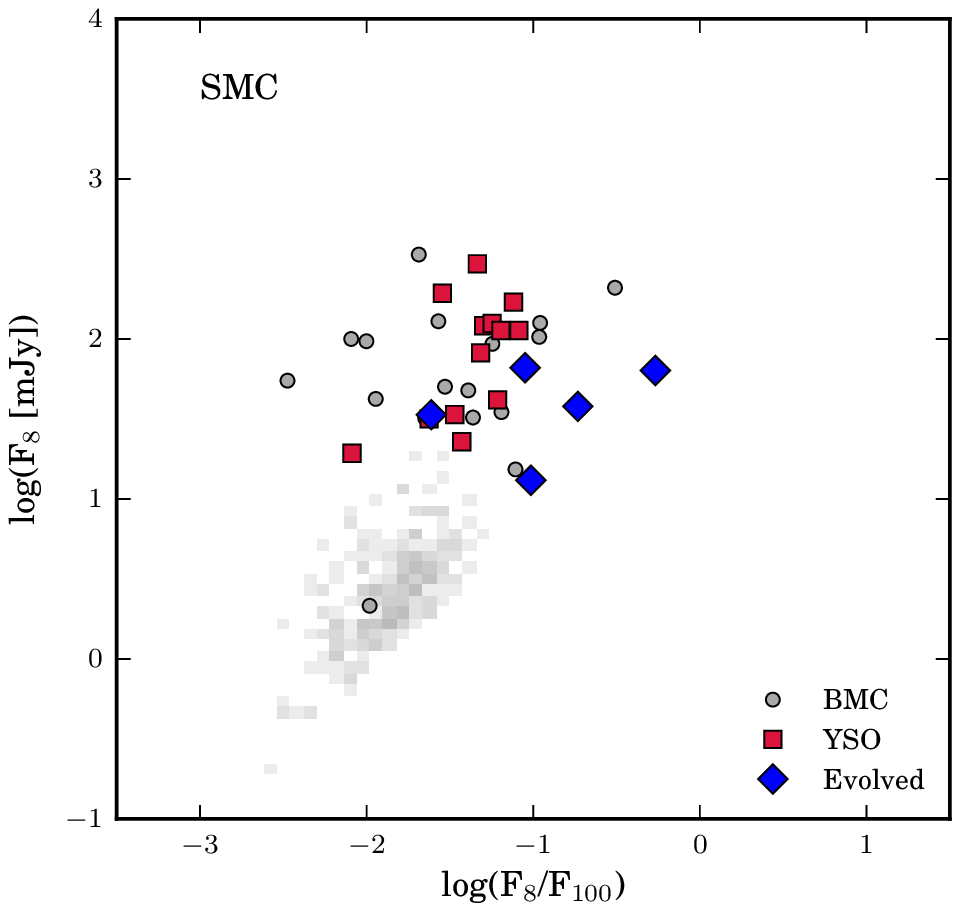}
 \caption[FIR CMDS]{IRAC $F_8$ and PACS $F_{100}$ CMD. Description as in Figure~\ref{fig:FIRCMD}.}
  \label{fig:FIRCMD_2}
\end{figure}

The evolved stars in our sample were identified from the HERITAGE far-IR BMC, as such they are subject to the same biases, selection effects and limitation discussed in detail by \cite{Seale2014}. Notably, the HERITAGE SPIRE observations are more sensitive than the PACS observations, thus faint sources near the sensitivity limit are detected only at 250 and 350 $\mu$m. This will adversely affect the detection of evolved stars which are expected to be brighter in the PACS bands compared to SPIRE.

In the HERITAGE images complex structures of knots and filaments due to ISM dust dominate the background emission; this prevents the detection of faint sources in regions of high background. While AGB stars are unlikely to be directly associated with dust clumps or filaments, many may lie along the same line of sight. This is even more problematic for massive stars which are preferentially found in higher ISM column densities. High background ($B_{250} > 10$ MJy sr$^{-1}$) regions cover 16\% of the LMC and 2\% of the SMC; in these regions the BMC is complete for sources brighter than 200 mJy  \citep{Seale2014}. This is a factor of two greater than the completeness limits in low surface brightness regions. If our sample is distributed evenly across the Magellanic Clouds, then bright extended emission in the Magellanic Clouds may mask around five bona fide evolved stars with far-IR emission greater than 100 mJy.  If we also take into account sources in the BMC which may potentially be evolved but were not included in our final list (as they are more likely to be YSOs) then this estimate increases to approximately 10 sources. The Magellanic Clouds should therefore host between 18 and 23  post-main sequence stars with far-IR fluxes greater than the BMC completeness limit, rather than the 13 we currently detect. The remaining 22 sources in our sample are faint and fall below this limit; here we do not have enough statistical information to provide a quantitative value for the total population of faint evolved stars with similar fluxes.

The identification of evolved stars in the BMC is biased towards brighter sources; these are prominent above the background and have an SED which is well sampled. To be included in our final list of probable evolved stars, each object must also be detected as a point-source in the mid-IR and be previously identified as an evolved star candidate in the literature or via photometric colour cuts. As such we do not discover any new objects in our vetted sample. Faint sources which are variable and optically bright (e.g. OGLE variable stars), but have a low mid-IR flux will be excluded from our list as they are difficult to characterize. To complicate matters further, faint sources have larger positional uncertainties in the far-IR and hence have a higher possibility of a photometric mismatch.

In Figures~\ref{fig:FIRCMD} and \ref{fig:FIRCMD_2} we compare our final sample of evolved stars to the entire HERITAGE BMC, the BMC candidate evolved stars which were identified as mismatch or probable YSOs and the BMC evolved star candidates confirmed spectroscopically as YSOs.  As noted earlier, the spectroscopically confirmed sources are limited to the brightest stars in our sample due to signal-to-noise ratio constraints. In the SPIRE $F_{250}$ and $F_{350}$ CMD there is a considerable amount of overlap between the different populations, and the sources are primarily separated based on their brightness at 250 $\mu$m. Here we see a clear distinction between the bright spectroscopically confirmed YSOs and the evolved stars at log($F_{250}$)$\sim$2.7, a similar cut is seen for the SMC despite its lower dust content which suggests a robust upper limit to the 250-micron flux of evolved stars. 

The majority of the sources that we conclusively identify fall well above the completeness limit of the {\em Herschel} BMC. Although we detect a moderate number of faint evolved sources below this line we have invariably missed others; it is difficult to put a number on how may more of these faint sources might be evolved. 
The small spread in the $F_{250}$/$F_{350}$ colour and the limited number of evolved stars detected in both bands prohibits the classification of sources using only their far-IR flux. If we are to clearly identify evolved sources additional mid-infrared data is required.  Figure~\ref{fig:FIRCMD_2} shows a CMD which combines mid- and far-IR photometry; here the various classes of objects tend to occupy different regions of CMD space although we are not able to completely separate the two populations. Furthermore, by combining far-IR data with 8 \mum photometry we effectively eliminate dust clumps and faint background galaxies from the CMD making it easier to distinguish the evolved sources.

\section{Summary and conclusions}  
\label{sec:conclusion}

In this paper, we present a catalog of thirty five evolved stars  and stellar end products which have been detected with PACS and SPIRE as part of the {\em Herschel} HERITAGE survey of the Magellanic Clouds. Our source identification method is based on mid-IR spectral confirmation, the SED characteristics, careful examination of the multiwavelength images and includes constraints on the luminosity and comparisons with YSO models.

The sources detected in our catalog cover a wide range in luminosity and hence initial mass. We have identified thirteen low- to intermediate mass post main-sequence stars, which are either currently losing-mass at high rates or are evolved post-superwind objects. We find ten high mass stars, including four of the fifteen known B[e] stars in the Magellanic Clouds, and three extreme RSGs which are highly enshrouded by dust.  Far-IR emission is also detected from the ejecta of SN1987A and from the supernova remnant N132D, both in the LMC. We have also identified far-IR emission from ten possible evolved star candidates. 

This catalog lays the groundwork for future studies of the far-IR emission from evolved stars and stellar end products in the Magellanic Clouds. In a future paper we will provide in-depth treatment of the far-IR emission via radiative transfer modeling to determine if the emission originates from the circumstellar envelope, or if it is due to swept-up ISM dust.

\vspace{0.4cm}

We thank the referee Jacco van Loon for their thorough and helpful comments. 
We acknowledge financial support from the NASA Herschel Science Center, JPL contracts \# 1381522, \# 1381650 \& \# 1350371.  Jones and Meixner acknowledge support from NASA grant, NNX14AN06G, for this work. Sargent acknowledges support from NASA grant NNX13AD54G.
This work is based on observations made with the {\em Herschel Space Observatory}, an ESA space observatory with science instruments provided by European-led Principal Investigator consortia and with important participation from NASA, and the {\em Spitzer} Space Telescope, which is operated by the Jet Propulsion Laboratory, California Institute of Technology under contract with NASA.
This research relied on the following resources: NASA's Astrophysics Data System and the SIMBAD and VizieR databases, operated at the Centre de Donn\'{e}es astronomiques de Strasbourg.

Facilities: {\em Herschel} (PACS, SPIRE) - European Space Agency's Herschel space observatory, {\em Spitzer} (IRS, IRAC, MIPS) - Spitzer Space Telescope satellite 




\def\aj{AJ}					
\def\actaa{Acta Astron.}                        
\def\araa{ARA\&A}				
\def\apj{ApJ}					
\def\apjl{ApJL}					
\def\apjs{ApJS}					
\def\ao{Appl.~Opt.}				
\def\apss{Ap\&SS}				
\def\aap{A\&A}					
\def\aapr{A\&A~Rev.}				
\def\aaps{A\&AS}				
\def\azh{AZh}					
\def\baas{BAAS}					
\def\jrasc{JRASC}				
\def\memras{MmRAS}				
\def\mnras{MNRAS}				
\def\pra{Phys.~Rev.~A}				
\def\prb{Phys.~Rev.~B}				
\def\prc{Phys.~Rev.~C}				
\def\prd{Phys.~Rev.~D}				
\def\pre{Phys.~Rev.~E}				
\def\prl{Phys.~Rev.~Lett.}			
\def\pasp{PASP}					
\def\pasj{PASJ}					
\def\qjras{QJRAS}				
\def\skytel{S\&T}				
\def\solphys{Sol.~Phys.}			
\def\sovast{Soviet~Ast.}			
\def\ssr{Space~Sci.~Rev.}			
\def\zap{ZAp}					
\def\nat{Nature}				
\def\iaucirc{IAU~Circ.}				
\def\aplett{Astrophys.~Lett.}			
\def\apspr{Astrophys.~Space~Phys.~Res.}		
\def\bain{Bull.~Astron.~Inst.~Netherlands}	
\def\fcp{Fund.~Cosmic~Phys.}			
\def\gca{Geochim.~Cosmochim.~Acta}		
\def\grl{Geophys.~Res.~Lett.}			
\def\jcp{J.~Chem.~Phys.}			
\def\jgr{J.~Geophys.~Res.}			
\def\jqsrt{J.~Quant.~Spec.~Radiat.~Transf.}	
\def\memsai{Mem.~Soc.~Astron.~Italiana}		
\def\nphysa{Nucl.~Phys.~A}			
\def\physrep{Phys.~Rep.}			
\def\physscr{Phys.~Scr}				
\def\planss{Planet.~Space~Sci.}			
\def\procspie{Proc.~SPIE}			
\def\icarus{Icarus}
\let\astap=\aap
\let\apjlett=\apjl
\let\apjsupp=\apjs
\let\applopt=\ao


\bibliographystyle{apj}


\begin{thebibliography}{}
\expandafter\ifx\csname natexlab\endcsname\relax\def\natexlab#1{#1}\fi

\bibitem[{{Aret} {et~al.}(2012){Aret}, {Kraus}, {Muratore}, \& {Borges
  Fernandes}}]{Aret2012}
{Aret}, A., {Kraus}, M., {Muratore}, M.~F., \& {Borges Fernandes}, M. 2012,
  \mnras, 423, 284

\bibitem[{{Barkat} \& {Wheeler}(1988)}]{Barkat1988}
{Barkat}, Z., \& {Wheeler}, J.~C. 1988, \apj, 332, 247

\bibitem[{{Barlow}(1987)}]{Barlow1987}
{Barlow}, M.~J. 1987, \mnras, 227, 161

\bibitem[{{Beelen} {et~al.}(2006){Beelen}, {Cox}, {Benford}, {Dowell},
  {Kov{\'a}cs}, {Bertoldi}, {Omont}, \& {Carilli}}]{Beelen2006}
{Beelen}, A., {Cox}, P., {Benford}, D.~J., {et~al.} 2006, \apj, 642, 694

\bibitem[{{Bernard-Salas} {et~al.}(2009){Bernard-Salas}, {Peeters}, {Sloan},
  {Gutenkunst}, {Matsuura}, {Tielens}, {Zijlstra}, \&
  {Houck}}]{BernardSalas2009}
{Bernard-Salas}, J., {Peeters}, E., {Sloan}, G.~C., {et~al.} 2009, \apj, 699,
  1541

\bibitem[{{Bernard-Salas} {et~al.}(2008){Bernard-Salas}, {Pottasch},
  {Gutenkunst}, {Morris}, \& {Houck}}]{BernardSalas2008}
{Bernard-Salas}, J., {Pottasch}, S.~R., {Gutenkunst}, S., {Morris}, P.~W., \&
  {Houck}, J.~R. 2008, \apj, 672, 274

\bibitem[{{Bertoldi} \& {Cox}(2002)}]{Bertoldi2002}
{Bertoldi}, F., \& {Cox}, P. 2002, \aap, 384, L11

\bibitem[{{Bica} {et~al.}(1999){Bica}, {Schmitt}, {Dutra}, \&
  {Oliveira}}]{Bica1999}
{Bica}, E.~L.~D., {Schmitt}, H.~R., {Dutra}, C.~M., \& {Oliveira}, H.~L. 1999,
  \aj, 117, 238

\bibitem[{{Bjorkman} {et~al.}(1998){Bjorkman}, {Miroshnichenko}, {Bjorkman},
  {Meade}, {Babler}, {Code}, {Anderson}, {Fox}, {Johnson}, {Weitenbeck},
  {Zellner}, \& {Lupie}}]{Bjorkman1998}
{Bjorkman}, K.~S., {Miroshnichenko}, A.~S., {Bjorkman}, J.~E., {et~al.} 1998,
  \apj, 509, 904

\bibitem[{{Blanco} {et~al.}(1980){Blanco}, {Blanco}, \&
  {McCarthy}}]{Blanco1980}
{Blanco}, V.~M., {Blanco}, B.~M., \& {McCarthy}, M.~F. 1980, \apj, 242, 938

\bibitem[{{Blum} {et~al.}(2006){Blum}, {Mould}, {Olsen}, {Frogel}, {Werner},
  {Meixner}, {Markwick-Kemper}, {Indebetouw}, {Whitney}, {Meade}, {Babler},
  {Churchwell}, {Gordon}, {Engelbracht}, {For}, {Misselt}, {Vijh}, {Leitherer},
  {Volk}, {Points}, {Reach}, {Hora}, {Bernard}, {Boulanger}, {Bracker},
  {Cohen}, {Fukui}, {Gallagher}, {Gorjian}, {Harris}, {Kelly}, {Kawamura},
  {Latter}, {Madden}, {Mizuno}, {Mizuno}, {Nota}, {Oey}, {Onishi}, {Paladini},
  {Panagia}, {Perez-Gonzalez}, {Shibai}, {Sato}, {Smith}, {Staveley-Smith},
  {Tielens}, {Ueta}, {Van Dyk}, \& {Zaritsky}}]{Blum2006}
{Blum}, R.~D., {Mould}, J.~R., {Olsen}, K.~A., {et~al.} 2006, \aj, 132, 2034

\bibitem[{{Boffi} \& {Stanghellini}(1994)}]{Boffi1994}
{Boffi}, F.~R., \& {Stanghellini}, L. 1994, \aap, 284, 248

\bibitem[{{Bonanos} {et~al.}(2009){Bonanos}, {Massa}, {Sewilo}, {Lennon},
  {Panagia}, {Smith}, {Meixner}, {Babler}, {Bracker}, {Meade}, {Gordon},
  {Hora}, {Indebetouw}, \& {Whitney}}]{Bonanos2009}
{Bonanos}, A.~Z., {Massa}, D.~L., {Sewilo}, M., {et~al.} 2009, \aj, 138, 1003

\bibitem[{{Bonanos} {et~al.}(2010){Bonanos}, {Lennon}, {K{\"o}hlinger}, {van
  Loon}, {Massa}, {Sewilo}, {Evans}, {Panagia}, {Babler}, {Block}, {Bracker},
  {Engelbracht}, {Gordon}, {Hora}, {Indebetouw}, {Meade}, {Meixner}, {Misselt},
  {Robitaille}, {Shiao}, \& {Whitney}}]{Bonanos2010}
{Bonanos}, A.~Z., {Lennon}, D.~J., {K{\"o}hlinger}, F., {et~al.} 2010, \aj,
  140, 416

\bibitem[{{Boroson} \& {Liebert}(1989)}]{Boroson1989}
{Boroson}, T.~A., \& {Liebert}, J. 1989, \apj, 339, 844

\bibitem[{{Boyer} {et~al.}(2010){Boyer}, {van Loon}, {McDonald}, {Gordon},
  {Babler}, {Block}, {Bracker}, {Engelbracht}, {Hora}, {Indebetouw}, {Meade},
  {Meixner}, {Misselt}, {Sewilo}, {Shiao}, \& {Whitney}}]{Boyer2010}
{Boyer}, M.~L., {van Loon}, J.~T., {McDonald}, I., {et~al.} 2010, \apjl, 711,
  L99

\bibitem[{{Boyer} {et~al.}(2011){Boyer}, {Srinivasan}, {van Loon}, {McDonald},
  {Meixner}, {Zaritsky}, {Gordon}, {Kemper}, {Babler}, {Block}, {Bracker},
  {Engelbracht}, {Hora}, {Indebetouw}, {Meade}, {Misselt}, {Robitaille},
  {Sewi{\l}o}, {Shiao}, \& {Whitney}}]{Boyer2011}
{Boyer}, M.~L., {Srinivasan}, S., {van Loon}, J.~T., {et~al.} 2011, \aj, 142,
  103

\bibitem[{{Boyer} {et~al.}(2012){Boyer}, {Srinivasan}, {Riebel}, {McDonald},
  {van Loon}, {Clayton}, {Gordon}, {Meixner}, {Sargent}, \&
  {Sloan}}]{Boyer2012}
{Boyer}, M.~L., {Srinivasan}, S., {Riebel}, D., {et~al.} 2012, \apj, 748, 40

\bibitem[{{Breysacher} {et~al.}(1999){Breysacher}, {Azzopardi}, \&
  {Testor}}]{Breysacher1999}
{Breysacher}, J., {Azzopardi}, M., \& {Testor}, G. 1999, \aaps, 137, 117

\bibitem[{{Buchanan} {et~al.}(2006){Buchanan}, {Kastner}, {Forrest}, {Hrivnak},
  {Sahai}, {Egan}, {Frank}, \& {Barnbaum}}]{Buchanan2006}
{Buchanan}, C.~L., {Kastner}, J.~H., {Forrest}, W.~J., {et~al.} 2006, \aj, 132,
  1890

\bibitem[{{Cioni} \& {Habing}(2003)}]{Cioni2003}
{Cioni}, M.-R.~L., \& {Habing}, H.~J. 2003, \aap, 402, 133

\bibitem[{{Cox} {et~al.}(2012){Cox}, {Kerschbaum}, {van Marle}, {Decin},
  {Ladjal}, {Mayer}, {Groenewegen}, {van Eck}, {Royer}, {Ottensamer}, {Ueta},
  {Jorissen}, {Mecina}, {Meliani}, {Luntzer}, {Blommaert}, {Posch},
  {Vandenbussche}, \& {Waelkens}}]{Cox2012}
{Cox}, N.~L.~J., {Kerschbaum}, F., {van Marle}, A.-J., {et~al.} 2012, \aap,
  537, A35

\bibitem[{{Crowther}(2007)}]{Crowther2007}
{Crowther}, P.~A. 2007, \araa, 45, 177

\bibitem[{{Cutri} \& {2MASS Team}(2004)}]{Cutri2004}
{Cutri}, R.~M., \& {2MASS Team}. 2004, in Bulletin of the American Astronomical
  Society, Vol.~36, American Astronomical Society Meeting Abstracts, 1487

\bibitem[{{de Grijp} {et~al.}(1987){de Grijp}, {Lub}, \& {Miley}}]{deGrijp1987}
{de Grijp}, M.~H.~K., {Lub}, J., \& {Miley}, G.~K. 1987, \aaps, 70, 95

\bibitem[{{Dijkstra} {et~al.}(2003){Dijkstra}, {Waters}, {Kemper}, {Min},
  {Matsuura}, {Zijlstra}, {de Koter}, \& {Dominik}}]{Dijkstra2003}
{Dijkstra}, C., {Waters}, L.~B.~F.~M., {Kemper}, F., {et~al.} 2003, \aap, 399,
  1037

\bibitem[{{Diolaiti} {et~al.}(2000){Diolaiti}, {Bendinelli}, {Bonaccini},
  {Close}, {Currie}, \& {Parmeggiani}}]{Diolaiti2000}
{Diolaiti}, E., {Bendinelli}, O., {Bonaccini}, D., {et~al.} 2000, in Society of
  Photo-Optical Instrumentation Engineers (SPIE) Conference Series, Vol. 4007,
  Adaptive Optical Systems Technology, ed. P.~L. {Wizinowich}, 879--888

\bibitem[{{Dwek} {et~al.}(2007){Dwek}, {Galliano}, \& {Jones}}]{Dwek2007}
{Dwek}, E., {Galliano}, F., \& {Jones}, A.~P. 2007, \apj, 662, 927

\bibitem[{{Egan} {et~al.}(2001){Egan}, {Van Dyk}, \& {Price}}]{Egan2001}
{Egan}, M.~P., {Van Dyk}, S.~D., \& {Price}, S.~D. 2001, \aj, 122, 1844

\bibitem[{{Elias} {et~al.}(1986){Elias}, {Frogel}, \& {Schwering}}]{Elias1986}
{Elias}, J.~H., {Frogel}, J.~A., \& {Schwering}, P.~B.~W. 1986, \apj, 302, 675

\bibitem[{{Fraser} {et~al.}(2008){Fraser}, {Hawley}, \& {Cook}}]{Fraser2008}
{Fraser}, O.~J., {Hawley}, S.~L., \& {Cook}, K.~H. 2008, \aj, 136, 1242

\bibitem[{{Gehrz}(1989)}]{Gehrz1989}
{Gehrz}, R. 1989, in IAU Symposium, Vol. 135, Interstellar Dust, ed.
  {L.~J.~Allamandola \& A.~G.~G.~M.~Tielens}, 445

\bibitem[{{Gordon} {et~al.}(2011){Gordon}, {Meixner}, {Meade}, {Whitney},
  {Engelbracht}, {Bot}, {Boyer}, {Lawton}, {Sewi{\l}o}, {Babler}, {Bernard},
  {Bracker}, {Block}, {Blum}, {Bolatto}, {Bonanos}, {Harris}, {Hora},
  {Indebetouw}, {Misselt}, {Reach}, {Shiao}, {Tielens}, {Carlson},
  {Churchwell}, {Clayton}, {Chen}, {Cohen}, {Fukui}, {Gorjian}, {Hony},
  {Israel}, {Kawamura}, {Kemper}, {Leroy}, {Li}, {Madden}, {Marble},
  {McDonald}, {Mizuno}, {Mizuno}, {Muller}, {Oliveira}, {Olsen}, {Onishi},
  {Paladini}, {Paradis}, {Points}, {Robitaille}, {Rubin}, {Sandstrom}, {Sato},
  {Shibai}, {Simon}, {Smith}, {Srinivasan}, {Vijh}, {Van Dyk}, {van Loon}, \&
  {Zaritsky}}]{Gordon2011}
{Gordon}, K.~D., {Meixner}, M., {Meade}, M.~R., {et~al.} 2011, \aj, 142, 102

\bibitem[{{Gouliermis} {et~al.}(2003){Gouliermis}, {Kontizas}, {Kontizas}, \&
  {Korakitis}}]{Gouliermis2003}
{Gouliermis}, D., {Kontizas}, M., {Kontizas}, E., \& {Korakitis}, R. 2003,
  \aap, 405, 111

\bibitem[{{Griffin} {et~al.}(2010){Griffin}, {Abergel}, {Abreu}, {Ade},
  {Andr{\'e}}, {Augueres}, {Babbedge}, {Bae}, {Baillie}, {Baluteau}, {Barlow},
  {Bendo}, {Benielli}, {Bock}, {Bonhomme}, {Brisbin}, {Brockley-Blatt},
  {Caldwell}, {Cara}, {Castro-Rodriguez}, {Cerulli}, {Chanial}, {Chen},
  {Clark}, {Clements}, {Clerc}, {Coker}, {Communal}, {Conversi}, {Cox},
  {Crumb}, {Cunningham}, {Daly}, {Davis}, {de Antoni}, {Delderfield}, {Devin},
  {di Giorgio}, {Didschuns}, {Dohlen}, {Donati}, {Dowell}, {Dowell}, {Duband},
  {Dumaye}, {Emery}, {Ferlet}, {Ferrand}, {Fontignie}, {Fox}, {Franceschini},
  {Frerking}, {Fulton}, {Garcia}, {Gastaud}, {Gear}, {Glenn}, {Goizel},
  {Griffin}, {Grundy}, {Guest}, {Guillemet}, {Hargrave}, {Harwit}, {Hastings},
  {Hatziminaoglou}, {Herman}, {Hinde}, {Hristov}, {Huang}, {Imhof}, {Isaak},
  {Israelsson}, {Ivison}, {Jennings}, {Kiernan}, {King}, {Lange}, {Latter},
  {Laurent}, {Laurent}, {Leeks}, {Lellouch}, {Levenson}, {Li}, {Li},
  {Lilienthal}, {Lim}, {Liu}, {Lu}, {Madden}, {Mainetti}, {Marliani}, {McKay},
  {Mercier}, {Molinari}, {Morris}, {Moseley}, {Mulder}, {Mur}, {Naylor},
  {Nguyen}, {O'Halloran}, {Oliver}, {Olofsson}, {Olofsson}, {Orfei}, {Page},
  {Pain}, {Panuzzo}, {Papageorgiou}, {Parks}, {Parr-Burman}, {Pearce},
  {Pearson}, {P{\'e}rez-Fournon}, {Pinsard}, {Pisano}, {Podosek}, {Pohlen},
  {Polehampton}, {Pouliquen}, {Rigopoulou}, {Rizzo}, {Roseboom}, {Roussel},
  {Rowan-Robinson}, {Rownd}, {Saraceno}, {Sauvage}, {Savage}, {Savini},
  {Sawyer}, {Scharmberg}, {Schmitt}, {Schneider}, {Schulz}, {Schwartz},
  {Shafer}, {Shupe}, {Sibthorpe}, {Sidher}, {Smith}, {Smith}, {Smith},
  {Spencer}, {Stobie}, {Sudiwala}, {Sukhatme}, {Surace}, {Stevens}, {Swinyard},
  {Trichas}, {Tourette}, {Triou}, {Tseng}, {Tucker}, {Turner}, {Vaccari},
  {Valtchanov}, {Vigroux}, {Virique}, {Voellmer}, {Walker}, {Ward}, {Waskett},
  {Weilert}, {Wesson}, {White}, {Whitehouse}, {Wilson}, {Winter}, {Woodcraft},
  {Wright}, {Xu}, {Zavagno}, {Zemcov}, {Zhang}, \& {Zonca}}]{Griffin2010}
{Griffin}, M.~J., {Abergel}, A., {Abreu}, A., {et~al.} 2010, \aap, 518, L3

\bibitem[{{Groenewegen} {et~al.}(2011){Groenewegen}, {Waelkens}, {Barlow},
  {Kerschbaum}, {Garcia-Lario}, {Cernicharo}, {Blommaert}, {Bouwman}, {Cohen},
  {Cox}, {Decin}, {Exter}, {Gear}, {Gomez}, {Hargrave}, {Henning},
  {Hutsem{\'e}kers}, {Ivison}, {Jorissen}, {Krause}, {Ladjal}, {Leeks}, {Lim},
  {Matsuura}, {Naz{\'e}}, {Olofsson}, {Ottensamer}, {Polehampton}, {Posch},
  {Rauw}, {Royer}, {Sibthorpe}, {Swinyard}, {Ueta}, {Vamvatira-Nakou},
  {Vandenbussche}, {van de Steene}, {van Eck}, {van Hoof}, {van Winckel},
  {Verdugo}, \& {Wesson}}]{Groenewegen2011}
{Groenewegen}, M.~A.~T., {Waelkens}, C., {Barlow}, M.~J., {et~al.} 2011, \aap,
  526, A162

\bibitem[{{Gruendl} \& {Chu}(2009)}]{Gruendl2009}
{Gruendl}, R.~A., \& {Chu}, Y.-H. 2009, \apjs, 184, 172

\bibitem[{{Gruendl} {et~al.}(2008){Gruendl}, {Chu}, {Seale}, {Matsuura},
  {Speck}, {Sloan}, \& {Looney}}]{Gruendl2008}
{Gruendl}, R.~A., {Chu}, Y.-H., {Seale}, J.~P., {et~al.} 2008, \apjl, 688, L9

\bibitem[{{Guha Niyogi} {et~al.}(2014){Guha Niyogi}, {Min}, {Meixner},
  {Waters}, {Seale}, \& {Tielens}}]{GuhaNiyogi2014}
{Guha Niyogi}, S., {Min}, M., {Meixner}, M., {et~al.} 2014, \aap, 569, A80

\bibitem[{{Harris} \& {Zaritsky}(2004)}]{Harris2004}
{Harris}, J., \& {Zaritsky}, D. 2004, \aj, 127, 1531

\bibitem[{{Henize}(1956)}]{Henize1956}
{Henize}, K.~G. 1956, \apjs, 2, 315

\bibitem[{{Herald} \& {Bianchi}(2007)}]{Herald2007}
{Herald}, J.~E., \& {Bianchi}, L. 2007, \apj, 661, 845

\bibitem[{{Heydari-Malayeri} \& {Melnick}(1992)}]{Heydari-Malayeri1992}
{Heydari-Malayeri}, M., \& {Melnick}, J. 1992, \aap, 258, L13

\bibitem[{{Hora} {et~al.}(2008){Hora}, {Cohen}, {Ellis}, {Meixner}, {Blum},
  {Latter}, {Whitney}, {Meade}, {Babler}, {Indebetouw}, {Gordon},
  {Engelbracht}, {For}, {Block}, {Misselt}, {Vijh}, \& {Leitherer}}]{Hora2008}
{Hora}, J.~L., {Cohen}, M., {Ellis}, R.~G., {et~al.} 2008, \aj, 135, 726

\bibitem[{{Humphreys} \& {Davidson}(1994)}]{Humphreys1994}
{Humphreys}, R.~M., \& {Davidson}, K. 1994, \pasp, 106, 1025

\bibitem[{{Hutsemekers}(1994)}]{Hutsemekers1994}
{Hutsemekers}, D. 1994, \aap, 281, L81

\bibitem[{{Indebetouw} {et~al.}(2014){Indebetouw}, {Matsuura}, {Dwek},
  {Zanardo}, {Barlow}, {Baes}, {Bouchet}, {Burrows}, {Chevalier}, {Clayton},
  {Fransson}, {Gaensler}, {Kirshner}, {Laki{\'c}evi{\'c}}, {Long}, {Lundqvist},
  {Mart{\'{\i}}-Vidal}, {Marcaide}, {McCray}, {Meixner}, {Ng}, {Park},
  {Sonneborn}, {Staveley-Smith}, {Vlahakis}, \& {van Loon}}]{Indebetouw2014}
{Indebetouw}, R., {Matsuura}, M., {Dwek}, E., {et~al.} 2014, \apjl, 782, L2

\bibitem[{{Ita} {et~al.}(2008){Ita}, {Onaka}, {Kato}, {Tanab{\'e}}, {Sakon},
  {Kaneda}, {Kawamura}, {Shimonishi}, {Wada}, {Usui}, {Koo}, {Matsuura},
  {Takahashi}, {Nakada}, {Hasegawa}, \& {Tamura}}]{Ita2008}
{Ita}, Y., {Onaka}, T., {Kato}, D., {et~al.} 2008, \pasj, 60, 435

\bibitem[{{Jones} {et~al.}(1994){Jones}, {Tielens}, {Hollenbach}, \&
  {McKee}}]{Jones1994}
{Jones}, A.~P., {Tielens}, A.~G.~G.~M., {Hollenbach}, D.~J., \& {McKee}, C.~F.
  1994, \apj, 433, 797

\bibitem[{{Jones} {et~al.}(2014){Jones}, {Kemper}, {Srinivasan}, {McDonald},
  {Sloan}, \& {Zijlstra}}]{Jones2014}
{Jones}, O.~C., {Kemper}, F., {Srinivasan}, S., {et~al.} 2014, \mnras, 440, 631

\bibitem[{{Jones} {et~al.}(2012){Jones}, {Kemper}, {Sargent}, {McDonald},
  {Gielen}, {Woods}, {Sloan}, {Boyer}, {Zijlstra}, {Clayton}, {Kraemer},
  {Srinivasan}, \& {Ruffle}}]{Jones2012}
{Jones}, O.~C., {Kemper}, F., {Sargent}, B.~A., {et~al.} 2012, \mnras, 427,
  3209

\bibitem[{{Joye} \& {Mandel}(2003)}]{Joye2003}
{Joye}, W.~A., \& {Mandel}, E. 2003, in Astronomical Society of the Pacific
  Conference Series, Vol. 295, Astronomical Data Analysis Software and Systems
  XII, ed. H.~E. {Payne}, R.~I. {Jedrzejewski}, \& R.~N. {Hook}, 489

\bibitem[{{Kastner} {et~al.}(2010){Kastner}, {Buchanan}, {Sahai}, {Forrest}, \&
  {Sargent}}]{Kastner2010}
{Kastner}, J.~H., {Buchanan}, C., {Sahai}, R., {Forrest}, W.~J., \& {Sargent},
  B.~A. 2010, \aj, 139, 1993

\bibitem[{{Kastner} {et~al.}(2006){Kastner}, {Buchanan}, {Sargent}, \&
  {Forrest}}]{Kastner2006}
{Kastner}, J.~H., {Buchanan}, C.~L., {Sargent}, B., \& {Forrest}, W.~J. 2006,
  \apjl, 638, L29

\bibitem[{{Kato} {et~al.}(2007){Kato}, {Nagashima}, {Nagayama}, {Kurita},
  {Koerwer}, {Kawai}, {Yamamuro}, {Zenno}, {Nishiyama}, {Baba}, {Kadowaki},
  {Haba}, {Hatano}, {Shimizu}, {Nishimura}, {Nagata}, {Sato}, {Murai},
  {Kawazu}, {Nakajima}, {Nakaya}, {Kandori}, {Kusakabe}, {Ishihara},
  {Kaneyasu}, {Hashimoto}, {Tamura}, {Tanab{\'e}}, {Ita}, {Matsunaga},
  {Nakada}, {Sugitani}, {Wakamatsu}, {Glass}, {Feast}, {Menzies}, {Whitelock},
  {Fourie}, {Stoffels}, {Evans}, \& {Hasegawa}}]{Kato2007}
{Kato}, D., {Nagashima}, C., {Nagayama}, T., {et~al.} 2007, \pasj, 59, 615

\bibitem[{{Keller} \& {Wood}(2006)}]{Keller2006}
{Keller}, S.~C., \& {Wood}, P.~R. 2006, \apj, 642, 834

\bibitem[{{Kemper} {et~al.}(2010){Kemper}, {Woods}, {Antoniou}, {Bernard},
  {Blum}, {Boyer}, {Chan}, {Chen}, {Cohen}, {Dijkstra}, {Engelbracht},
  {Galametz}, {Galliano}, {Gielen}, {Gordon}, {Gorjian}, {Harris}, {Hony},
  {Hora}, {Indebetouw}, {Jones}, {Kawamura}, {Lagadec}, {Lawton}, {Leisenring},
  {Madden}, {Marengo}, {Matsuura}, {McDonald}, {McGuire}, {Meixner}, {Mulia},
  {O'Halloran}, {Oliveira}, {Paladini}, {Paradis}, {Reach}, {Rubin},
  {Sandstrom}, {Sargent}, {Sewilo}, {Shiao}, {Sloan}, {Speck}, {Srinivasan},
  {Szczerba}, {Tielens}, {van Aarle}, {Van Dyk}, {van Loon}, {Van Winckel},
  {Vijh}, {Volk}, {Whitney}, {Wilkins}, \& {Zijlstra}}]{Kemper2010}
{Kemper}, F., {Woods}, P.~M., {Antoniou}, V., {et~al.} 2010, \pasp, 122, 683

\bibitem[{{Kerschbaum} {et~al.}(2010){Kerschbaum}, {Ladjal}, {Ottensamer},
  {Groenewegen}, {Mecina}, {Blommaert}, {Baumann}, {Decin}, {Vandenbussche},
  {Waelkens}, {Posch}, {Huygen}, {De Meester}, {Regibo}, {Royer}, {Exter}, \&
  {Jean}}]{Kerschbaum2010}
{Kerschbaum}, F., {Ladjal}, D., {Ottensamer}, R., {et~al.} 2010, \aap, 518,
  L140

\bibitem[{{Kim} {et~al.}(2010){Kim}, {Kwon}, {Madden}, {Meixner}, {Hony},
  {Panuzzo}, {Sauvage}, {Roman-Duval}, {Gordon}, {Engelbracht}, {Israel},
  {Misselt}, {Okumura}, {Li}, {Bolatto}, {Skibba}, {Galliano}, {Matsuura},
  {Bernard}, {Bot}, {Galametz}, {Hughes}, {Kawamura}, {Onishi}, {Paradis},
  {Poglitsch}, {Reach}, {Robitaille}, {Rubio}, \& {Tielens}}]{Kim2010}
{Kim}, S., {Kwon}, E., {Madden}, S.~C., {et~al.} 2010, \aap, 518, L75

\bibitem[{{Koz{\l}owski} \& {Kochanek}(2009)}]{Kozlowski2009}
{Koz{\l}owski}, S., \& {Kochanek}, C.~S. 2009, \apj, 701, 508

\bibitem[{{Kraus} {et~al.}(2007){Kraus}, {Borges Fernandes}, \& {de
  Ara{\'u}jo}}]{Kraus2007}
{Kraus}, M., {Borges Fernandes}, M., \& {de Ara{\'u}jo}, F.~X. 2007, \aap, 463,
  627

\bibitem[{{Laki{\'c}evi{\'c}} {et~al.}(2015){Laki{\'c}evi{\'c}}, {van Loon},
  {Meixner}, {Gordon}, {Bot}, {Roman-Duval}, {Babler}, {Bolatto},
  {Engelbracht}, {Filipovi{\'c}}, {Hony}, {Indebetouw}, {Misselt}, {Montiel},
  {Okumura}, {Panuzzo}, {Patat}, {Sauvage}, {Seale}, {Sonneborn}, {Temim},
  {Uro{\v s}evi{\'c}}, \& {Zanardo}}]{Lakicevic2015}
{Laki{\'c}evi{\'c}}, M., {van Loon}, J.~T., {Meixner}, M., {et~al.} 2015, \apj,
  799, 50

\bibitem[{{Lamers} {et~al.}(1998){Lamers}, {Bastiaanse}, {Aerts}, \&
  {Spoon}}]{Lamers1998b}
{Lamers}, H.~J.~G.~L.~M., {Bastiaanse}, M.~V., {Aerts}, C., \& {Spoon},
  H.~W.~W. 1998, \aap, 335, 605

\bibitem[{{Lennon} {et~al.}(1993){Lennon}, {Wobig}, {Kudritzki}, \&
  {Stahl}}]{Lennon1993}
{Lennon}, D.~J., {Wobig}, D., {Kudritzki}, R.-P., \& {Stahl}, O. 1993, \ssr,
  66, 207

\bibitem[{{Levesque} {et~al.}(2007){Levesque}, {Massey}, {Olsen}, \&
  {Plez}}]{Levesque2007}
{Levesque}, E.~M., {Massey}, P., {Olsen}, K.~A.~G., \& {Plez}, B. 2007, \apj,
  667, 202

\bibitem[{{Lindsay}(1961)}]{Lindsay1961}
{Lindsay}, E.~M. 1961, \aj, 66, 169

\bibitem[{{Maercker} {et~al.}(2012){Maercker}, {Mohamed}, {Vlemmings},
  {Ramstedt}, {Groenewegen}, {Humphreys}, {Kerschbaum}, {Lindqvist},
  {Olofsson}, {Paladini}, {Wittkowski}, {de Gregorio-Monsalvo}, \&
  {Nyman}}]{Maercker2012}
{Maercker}, M., {Mohamed}, S., {Vlemmings}, W.~H.~T., {et~al.} 2012, \nat, 490,
  232

\bibitem[{{Marshall} {et~al.}(2004){Marshall}, {van Loon}, {Matsuura}, {Wood},
  {Zijlstra}, \& {Whitelock}}]{Marshall2004}
{Marshall}, J.~R., {van Loon}, J.~T., {Matsuura}, M., {et~al.} 2004, \mnras,
  355, 1348

\bibitem[{{Massey}(2002)}]{Massey2002}
{Massey}, P. 2002, \apjs, 141, 81

\bibitem[{{Massey} \& {Olsen}(2003)}]{Massey2003}
{Massey}, P., \& {Olsen}, K.~A.~G. 2003, \aj, 126, 2867

\bibitem[{{Matsuura} {et~al.}(2009){Matsuura}, {Barlow}, {Zijlstra},
  {Whitelock}, {Cioni}, {Groenewegen}, {Volk}, {Kemper}, {Kodama}, {Lagadec},
  {Meixner}, {Sloan}, \& {Srinivasan}}]{Matsuura2009}
{Matsuura}, M., {Barlow}, M.~J., {Zijlstra}, A.~A., {et~al.} 2009, \mnras, 396,
  918

\bibitem[{{Matsuura} {et~al.}(2011){Matsuura}, {Dwek}, {Meixner}, {Otsuka},
  {Babler}, {Barlow}, {Roman-Duval}, {Engelbracht}, {Sandstrom},
  {Laki{\'c}evi{\'c}}, {van Loon}, {Sonneborn}, {Clayton}, {Long}, {Lundqvist},
  {Nozawa}, {Gordon}, {Hony}, {Panuzzo}, {Okumura}, {Misselt}, {Montiel}, \&
  {Sauvage}}]{Matsuura2011}
{Matsuura}, M., {Dwek}, E., {Meixner}, M., {et~al.} 2011, Science, 333, 1258

\bibitem[{{Matsuura} {et~al.}(2014){Matsuura}, {Bernard-Salas}, {Lloyd Evans},
  {Volk}, {Hrivnak}, {Sloan}, {Chu}, {Gruendl}, {Kraemer}, {Peeters},
  {Szczerba}, {Wood}, {Zijlstra}, {Hony}, {Ita}, {Kamath}, {Lagadec}, {Parker},
  {Reid}, {Shimonishi}, {Van Winckel}, {Woods}, {Kemper}, {Meixner}, {Otsuka},
  {Sahai}, {Sargent}, {Hora}, \& {McDonald}}]{Matsuura2014}
{Matsuura}, M., {Bernard-Salas}, J., {Lloyd Evans}, T., {et~al.} 2014, \mnras,
  439, 1472

\bibitem[{{Matsuura} {et~al.}(2015){Matsuura}, {Dwek}, {Barlow}, {Babler},
  {Baes}, {Meixner}, {Cernicharo}, {Clayton}, {Dunne}, {Fransson}, {Fritz},
  {Gear}, {Gomez}, {Groenewegen}, {Indebetouw}, {Ivison}, {Jerkstrand},
  {Lebouteiller}, {Lim}, {Lundqvist}, {Pearson}, {Roman-Duval}, {Royer},
  {Staveley-Smith}, {Swinyard}, {van Hoof}, {van Loon}, {Verstappen}, {Wesson},
  {Zanardo}, {Blommaert}, {Decin}, {Reach}, {Sonneborn}, {Van de Steene}, \&
  {Yates}}]{Matsuura2015}
{Matsuura}, M., {Dwek}, E., {Barlow}, M.~J., {et~al.} 2015, \apj, 800, 50

\bibitem[{{Meixner} {et~al.}(2006){Meixner}, {Gordon}, {Indebetouw}, {Hora},
  {Whitney}, {Blum}, {Reach}, {Bernard}, {Meade}, {Babler}, {Engelbracht},
  {For}, {Misselt}, {Vijh}, {Leitherer}, {Cohen}, {Churchwell}, {Boulanger},
  {Frogel}, {Fukui}, {Gallagher}, {Gorjian}, {Harris}, {Kelly}, {Kawamura},
  {Kim}, {Latter}, {Madden}, {Markwick-Kemper}, {Mizuno}, {Mizuno}, {Mould},
  {Nota}, {Oey}, {Olsen}, {Onishi}, {Paladini}, {Panagia}, {Perez-Gonzalez},
  {Shibai}, {Sato}, {Smith}, {Staveley-Smith}, {Tielens}, {Ueta}, {van Dyk},
  {Volk}, {Werner}, \& {Zaritsky}}]{Meixner2006}
{Meixner}, M., {Gordon}, K.~D., {Indebetouw}, R., {et~al.} 2006, \aj, 132, 2268

\bibitem[{{Meixner} {et~al.}(2010){Meixner}, {Hony}, {Madden}, {Gordon},
  {Engelbracht}, {Tielens}, {Bolatto}, {Duval}, {Babler}, {Misselt},
  {Galliano}, {Fukui}, {Indebetouw}, {Israel}, {Sewilo}, {Kawamura}, {Kim},
  {Long}, {Kemper}, {Bernard}, {Okumura}, {Panuzzo}, {Poglitsch}, {Sauvage},
  {Srinivasan}, {Robitaille}, {van Loon}, {Whitney}, {Oliveira}, {Hora}, {Bot},
  {Skibba}, \& {HERITAGE Team}}]{Meixner2010AAS}
{Meixner}, M., {Hony}, S., {Madden}, S., {et~al.} 2010, in Bulletin of the
  American Astronomical Society, Vol.~42, American Astronomical Society Meeting
  Abstracts 215, 459.22

\bibitem[{{Meixner} {et~al.}(2013){Meixner}, {Panuzzo}, {Roman-Duval},
  {Engelbracht}, {Babler}, {Seale}, {Hony}, {Montiel}, {Sauvage}, {Gordon},
  {Misselt}, {Okumura}, {Chanial}, {Beck}, {Bernard}, {Bolatto}, {Bot},
  {Boyer}, {Carlson}, {Clayton}, {Chen}, {Cormier}, {Fukui}, {Galametz},
  {Galliano}, {Hora}, {Hughes}, {Indebetouw}, {Israel}, {Kawamura}, {Kemper},
  {Kim}, {Kwon}, {Lebouteiller}, {Li}, {Long}, {Madden}, {Matsuura}, {Muller},
  {Oliveira}, {Onishi}, {Otsuka}, {Paradis}, {Poglitsch}, {Reach},
  {Robitaille}, {Rubio}, {Sargent}, {Sewi{\l}o}, {Skibba}, {Smith},
  {Srinivasan}, {Tielens}, {van Loon}, \& {Whitney}}]{Meixner2013}
{Meixner}, M., {Panuzzo}, P., {Roman-Duval}, J., {et~al.} 2013, \aj, 146, 62

\bibitem[{{Mennickent} {et~al.}(2002){Mennickent}, {Pietrzy{\'n}ski}, {Gieren},
  \& {Szewczyk}}]{Mennickent2002}
{Mennickent}, R.~E., {Pietrzy{\'n}ski}, G., {Gieren}, W., \& {Szewczyk}, O.
  2002, \aap, 393, 887

\bibitem[{{Meynadier} \& {Heydari-Malayeri}(2007)}]{Meynadier2007}
{Meynadier}, F., \& {Heydari-Malayeri}, M. 2007, \aap, 461, 565

\bibitem[{{Meyssonnier} \& {Azzopardi}(1993)}]{Meyssonnier1993}
{Meyssonnier}, N., \& {Azzopardi}, M. 1993, \aaps, 102, 451

\bibitem[{{Micha{\l}owski} {et~al.}(2010){Micha{\l}owski}, {Murphy}, {Hjorth},
  {Watson}, {Gall}, \& {Dunlop}}]{Michalowski2010}
{Micha{\l}owski}, M.~J., {Murphy}, E.~J., {Hjorth}, J., {et~al.} 2010, \aap,
  522, A15

\bibitem[{{Moffat}(1991)}]{Moffat1991}
{Moffat}, A.~F.~J. 1991, \aap, 244, L9

\bibitem[{{Morse} {et~al.}(1995){Morse}, {Winkler}, \& {Kirshner}}]{Morse1995}
{Morse}, J.~A., {Winkler}, P.~F., \& {Kirshner}, R.~P. 1995, \aj, 109, 2104

\bibitem[{{Murphy} \& {Bessell}(2000)}]{Murphy2000}
{Murphy}, M.~T., \& {Bessell}, M.~S. 2000, \mnras, 311, 741

\bibitem[{{Ngeow} \& {Kanbur}(2008)}]{Ngeow2008}
{Ngeow}, C., \& {Kanbur}, S.~M. 2008, \apj, 679, 76

\bibitem[{{Ohnaka} {et~al.}(2008){Ohnaka}, {Driebe}, {Hofmann}, {Weigelt}, \&
  {Wittkowski}}]{Ohnaka2008}
{Ohnaka}, K., {Driebe}, T., {Hofmann}, K.-H., {Weigelt}, G., \& {Wittkowski},
  M. 2008, \aap, 484, 371

\bibitem[{{Oliveira} {et~al.}(2009){Oliveira}, {van Loon}, {Chen}, {Tielens},
  {Sloan}, {Woods}, {Kemper}, {Indebetouw}, {Gordon}, {Boyer}, {Shiao},
  {Madden}, {Speck}, {Meixner}, \& {Marengo}}]{Oliveira2009}
{Oliveira}, J.~M., {van Loon}, J.~T., {Chen}, C.-H.~R., {et~al.} 2009, \apj,
  707, 1269

\bibitem[{{Oliveira} {et~al.}(2013){Oliveira}, {van Loon}, {Sloan},
  {Sewi{\l}o}, {Kraemer}, {Wood}, {Indebetouw}, {Filipovi{\'c}}, {Crawford},
  {Wong}, {Hora}, {Meixner}, {Robitaille}, {Shiao}, \& {Simon}}]{Oliveira2013}
{Oliveira}, J.~M., {van Loon}, J.~T., {Sloan}, G.~C., {et~al.} 2013, \mnras,
  428, 3001

\bibitem[{{Otsuka} {et~al.}(2010){Otsuka}, {van Loon}, {Long}, {Meixner},
  {Matsuura}, {Reach}, {Roman-Duval}, {Gordon}, {Sauvage}, {Hony}, {Misselt},
  {Engelbracht}, {Panuzzo}, {Okumura}, {Woods}, {Kemper}, \&
  {Sloan}}]{Otsuka2010}
{Otsuka}, M., {van Loon}, J.~T., {Long}, K.~S., {et~al.} 2010, \aap, 518, L139

\bibitem[{{Pilbratt}(2010)}]{Pilbratt2010}
{Pilbratt}, G. 2010, in Bulletin of the American Astronomical Society, Vol.~41,
  Bulletin of the American Astronomical Society, 418

\bibitem[{{Poglitsch} {et~al.}(2010){Poglitsch}, {Waelkens}, {Geis},
  {Feuchtgruber}, {Vandenbussche}, {Rodriguez}, {Krause}, {Renotte}, {van
  Hoof}, {Saraceno}, {Cepa}, {Kerschbaum}, {Agn{\`e}se}, {Ali}, {Altieri},
  {Andreani}, {Augueres}, {Balog}, {Barl}, {Bauer}, {Belbachir}, {Benedettini},
  {Billot}, {Boulade}, {Bischof}, {Blommaert}, {Callut}, {Cara}, {Cerulli},
  {Cesarsky}, {Contursi}, {Creten}, {De Meester}, {Doublier}, {Doumayrou},
  {Duband}, {Exter}, {Genzel}, {Gillis}, {Gr{\"o}zinger}, {Henning},
  {Herreros}, {Huygen}, {Inguscio}, {Jakob}, {Jamar}, {Jean}, {de Jong},
  {Katterloher}, {Kiss}, {Klaas}, {Lemke}, {Lutz}, {Madden}, {Marquet},
  {Martignac}, {Mazy}, {Merken}, {Montfort}, {Morbidelli}, {M{\"u}ller},
  {Nielbock}, {Okumura}, {Orfei}, {Ottensamer}, {Pezzuto}, {Popesso},
  {Putzeys}, {Regibo}, {Reveret}, {Royer}, {Sauvage}, {Schreiber}, {Stegmaier},
  {Schmitt}, {Schubert}, {Sturm}, {Thiel}, {Tofani}, {Vavrek}, {Wetzstein},
  {Wieprecht}, \& {Wiezorrek}}]{Poglitsch2010}
{Poglitsch}, A., {Waelkens}, C., {Geis}, N., {et~al.} 2010, \aap, 518, L2

\bibitem[{{Porter}(2003)}]{Porter2003}
{Porter}, J.~M. 2003, \aap, 398, 631

\bibitem[{{Reid} \& {Parker}(2006)}]{Reid2006}
{Reid}, W.~A., \& {Parker}, Q.~A. 2006, \mnras, 373, 521

\bibitem[{{Reid} \& {Parker}(2012)}]{Reid2012}
---. 2012, \mnras, 425, 355

\bibitem[{{Rho} {et~al.}(2009){Rho}, {Reach}, {Tappe}, {Hwang}, {Slavin},
  {Kozasa}, \& {Dunne}}]{Rho2009}
{Rho}, J., {Reach}, W.~T., {Tappe}, A., {et~al.} 2009, \apj, 700, 579

\bibitem[{{Riebel} {et~al.}(2012){Riebel}, {Srinivasan}, {Sargent}, \&
  {Meixner}}]{Riebel2012}
{Riebel}, D., {Srinivasan}, S., {Sargent}, B., \& {Meixner}, M. 2012, \apj,
  753, 71

\bibitem[{{Robitaille} {et~al.}(2007){Robitaille}, {Whitney}, {Indebetouw}, \&
  {Wood}}]{Robitaille2007}
{Robitaille}, T.~P., {Whitney}, B.~A., {Indebetouw}, R., \& {Wood}, K. 2007,
  \apjs, 169, 328

\bibitem[{{Robitaille} {et~al.}(2006){Robitaille}, {Whitney}, {Indebetouw},
  {Wood}, \& {Denzmore}}]{Robitaille2006}
{Robitaille}, T.~P., {Whitney}, B.~A., {Indebetouw}, R., {Wood}, K., \&
  {Denzmore}, P. 2006, \apjs, 167, 256

\bibitem[{{Roche} {et~al.}(1993){Roche}, {Aitken}, \& {Smith}}]{Roche1993}
{Roche}, P.~F., {Aitken}, D.~K., \& {Smith}, C.~H. 1993, \mnras, 262, 301

\bibitem[{{Ruffle} {et~al.}(2015){Ruffle}, {Kemper}, {Jones}, {Sloan},
  {Kraemer}, {Woods}, {Boyer}, {Srinivasan}, {Antoniou}, {Lagadec}, {Matsuura},
  {McDonald}, {Oliveira}, {Sargent}, {Sewi{\l}o}, {Szczerba}, {van Loon},
  {Volk}, \& {Zijlstra}}]{Ruffle2015}
{Ruffle}, P.~M.~E., {Kemper}, F., {Jones}, O.~C., {et~al.} 2015, \mnras, 451,
  3504

\bibitem[{{Russell} \& {Dopita}(1992)}]{Russell1992}
{Russell}, S.~C., \& {Dopita}, M.~A. 1992, \apj, 384, 508

\bibitem[{{Samus} {et~al.}(2004){Samus}, {Durlevich}, \& {et al.}}]{Samus2004}
{Samus}, N.~N., {Durlevich}, O.~V., \& {et al.} 2004, VizieR Online Data
  Catalog, 2250, 0

\bibitem[{{Schlegel} {et~al.}(1998){Schlegel}, {Finkbeiner}, \&
  {Davis}}]{Schlegel1998}
{Schlegel}, D.~J., {Finkbeiner}, D.~P., \& {Davis}, M. 1998, \apj, 500, 525

\bibitem[{{Seale} {et~al.}(2014){Seale}, {Meixner}, {Sewi{\l}o}, {Babler},
  {Engelbracht}, {Gordon}, {Hony}, {Misselt}, {Montiel}, {Okumura}, {Panuzzo},
  {Roman-Duval}, {Sauvage}, {Boyer}, {Chen}, {Indebetouw}, {Matsuura},
  {Oliveira}, {Srinivasan}, {van Loon}, {Whitney}, \& {Woods}}]{Seale2014}
{Seale}, J.~P., {Meixner}, M., {Sewi{\l}o}, M., {et~al.} 2014, \aj, 148, 124

\bibitem[{{Sewi{\l}o} {et~al.}(2013){Sewi{\l}o}, {Carlson}, {Seale},
  {Indebetouw}, {Meixner}, {Whitney}, {Robitaille}, {Oliveira}, {Gordon},
  {Meade}, {Babler}, {Hora}, {Block}, {Misselt}, {van Loon}, {Chen},
  {Churchwell}, \& {Shiao}}]{Sewilo2013}
{Sewi{\l}o}, M., {Carlson}, L.~R., {Seale}, J.~P., {et~al.} 2013, \apj, 778, 15

\bibitem[{{Shaw} {et~al.}(2001){Shaw}, {Stanghellini}, {Mutchler}, {Balick}, \&
  {Blades}}]{Shaw2001}
{Shaw}, R.~A., {Stanghellini}, L., {Mutchler}, M., {Balick}, B., \& {Blades},
  J.~C. 2001, \apj, 548, 727

\bibitem[{{Shaw} {et~al.}(2006){Shaw}, {Stanghellini}, {Villaver}, \&
  {Mutchler}}]{Shaw2006}
{Shaw}, R.~A., {Stanghellini}, L., {Villaver}, E., \& {Mutchler}, M. 2006,
  \apjs, 167, 201

\bibitem[{{Sheets} {et~al.}(2013){Sheets}, {Bolatto}, {van Loon}, {Sandstrom},
  {Simon}, {Oliveira}, \& {Barb{\'a}}}]{Sheets2013}
{Sheets}, H.~A., {Bolatto}, A.~D., {van Loon}, J.~T., {et~al.} 2013, \apj, 771,
  111

\bibitem[{{Skrutskie} {et~al.}(2006){Skrutskie}, {Cutri}, {Stiening},
  {Weinberg}, {Schneider}, {Carpenter}, {Beichman}, {Capps}, {Chester},
  {Elias}, {Huchra}, {Liebert}, {Lonsdale}, {Monet}, {Price}, {Seitzer},
  {Jarrett}, {Kirkpatrick}, {Gizis}, {Howard}, {Evans}, {Fowler}, {Fullmer},
  {Hurt}, {Light}, {Kopan}, {Marsh}, {McCallon}, {Tam}, {Van Dyk}, \&
  {Wheelock}}]{Skrutskie2006}
{Skrutskie}, M.~F., {Cutri}, R.~M., {Stiening}, R., {et~al.} 2006, \aj, 131,
  1163

\bibitem[{{Sloan} {et~al.}(2014){Sloan}, {Lagadec}, {Zijlstra}, {Kraemer},
  {Weis}, {Matsuura}, {Volk}, {Peeters}, {Duley}, {Cami}, {Bernard-Salas},
  {Kemper}, \& {Sahai}}]{Sloan2014}
{Sloan}, G.~C., {Lagadec}, E., {Zijlstra}, A.~A., {et~al.} 2014, \apj, 791, 28

\bibitem[{{Smith} \& {MCELS Team}(1998)}]{MCELS1998}
{Smith}, R.~C., \& {MCELS Team}. 1998, \pasa, 15, 163

\bibitem[{{Srinivasan} {et~al.}(2009){Srinivasan}, {Meixner}, {Leitherer},
  {Vijh}, {Volk}, {Blum}, {Babler}, {Block}, {Bracker}, {Cohen}, {Engelbracht},
  {For}, {Gordon}, {Harris}, {Hora}, {Indebetouw}, {Markwick-Kemper}, {Meade},
  {Misselt}, {Sewilo}, \& {Whitney}}]{Srinivasan2009}
{Srinivasan}, S., {Meixner}, M., {Leitherer}, C., {et~al.} 2009, \aj, 137, 4810

\bibitem[{{Stahl} {et~al.}(1983){Stahl}, {Wolf}, {Klare}, {Cassatella},
  {Krautter}, {Persi}, \& {Ferrari-Toniolo}}]{Stahl1983}
{Stahl}, O., {Wolf}, B., {Klare}, G., {et~al.} 1983, \aap, 127, 49

\bibitem[{{Stanghellini} {et~al.}(1999){Stanghellini}, {Blades}, {Osmer},
  {Barlow}, \& {Liu}}]{Stanghellini1999}
{Stanghellini}, L., {Blades}, J.~C., {Osmer}, S.~J., {Barlow}, M.~J., \& {Liu},
  X.-W. 1999, \apj, 510, 687

\bibitem[{{Stanghellini} {et~al.}(2007){Stanghellini}, {Garc{\'{\i}}a-Lario},
  {Garc{\'{\i}}a-Hern{\'a}ndez}, {Perea-Calder{\'o}n}, {Davies}, {Manchado},
  {Villaver}, \& {Shaw}}]{Stanghellini2007}
{Stanghellini}, L., {Garc{\'{\i}}a-Lario}, P., {Garc{\'{\i}}a-Hern{\'a}ndez},
  D.~A., {et~al.} 2007, \apj, 671, 1669

\bibitem[{{Stanghellini} {et~al.}(2003){Stanghellini}, {Shaw}, {Balick},
  {Mutchler}, {Blades}, \& {Villaver}}]{Stanghellini2003}
{Stanghellini}, L., {Shaw}, R.~A., {Balick}, B., {et~al.} 2003, \apj, 596, 997

\bibitem[{{Stanghellini} {et~al.}(2005){Stanghellini}, {Shaw}, \&
  {Gilmore}}]{Stanghellini2005}
{Stanghellini}, L., {Shaw}, R.~A., \& {Gilmore}, D. 2005, \apj, 622, 294

\bibitem[{{Szczygie{\l}} {et~al.}(2010){Szczygie{\l}}, {Stanek}, {Bonanos},
  {Pojma{\'n}ski}, {Pilecki}, \& {Prieto}}]{Szczygiel2010}
{Szczygie{\l}}, D.~M., {Stanek}, K.~Z., {Bonanos}, A.~Z., {et~al.} 2010, \aj,
  140, 14

\bibitem[{{Szewczyk} {et~al.}(2009){Szewczyk}, {Pietrzy{\'n}ski}, {Gieren},
  {Ciechanowska}, {Bresolin}, \& {Kudritzki}}]{Szewczyk2009}
{Szewczyk}, O., {Pietrzy{\'n}ski}, G., {Gieren}, W., {et~al.} 2009, \aj, 138,
  1661

\bibitem[{{Tappe} {et~al.}(2006){Tappe}, {Rho}, \& {Reach}}]{Tappe2006}
{Tappe}, A., {Rho}, J., \& {Reach}, W.~T. 2006, \apj, 653, 267

\bibitem[{{Tchernyshyov} {et~al.}(2015){Tchernyshyov}, {Meixner}, {Seale},
  {Fox}, {Friedman}, {Dwek}, {Galliano}, \& {Sembach}}]{Tchernyshyov2015}
{Tchernyshyov}, K., {Meixner}, M., {Seale}, J., {et~al.} 2015, ArXiv e-prints,
  arXiv:1503.08852

\bibitem[{{Temim} {et~al.}(2015){Temim}, {Dwek}, {Tchernyshyov}, {Boyer},
  {Meixner}, {Gall}, \& {Roman-Duval}}]{Temim2015}
{Temim}, T., {Dwek}, E., {Tchernyshyov}, K., {et~al.} 2015, \apj, 799, 158

\bibitem[{{Testor} {et~al.}(2014){Testor}, {Heydari-Malayeri}, {Chen},
  {Lemaire}, {Sewi{\l}o}, \& {Diana}}]{Testor2014}
{Testor}, G., {Heydari-Malayeri}, M., {Chen}, C.-H.~R., {et~al.} 2014, \aap,
  564, A31

\bibitem[{{Trams} {et~al.}(1999){Trams}, {van Loon}, {Waters}, {Zijlstra},
  {Loup}, {Whitelock}, {Groenewegen}, {Blommaert}, {Siebenmorgen}, {Heske}, \&
  {Feast}}]{Trams1999}
{Trams}, N.~R., {van Loon}, J.~T., {Waters}, L.~B.~F.~M., {et~al.} 1999, \aap,
  346, 843

\bibitem[{{van Aarle} {et~al.}(2011){van Aarle}, {van Winckel}, {Lloyd Evans},
  {Ueta}, {Wood}, \& {Ginsburg}}]{vanAarle2011}
{van Aarle}, E., {van Winckel}, H., {Lloyd Evans}, T., {et~al.} 2011, \aap,
  530, A90

\bibitem[{{van der Marel} \& {Cioni}(2001)}]{vanderMarel2001}
{van der Marel}, R.~P., \& {Cioni}, M.-R.~L. 2001, \aj, 122, 1807

\bibitem[{{van Loon} {et~al.}(2005{\natexlab{a}}){van Loon}, {Cioni},
  {Zijlstra}, \& {Loup}}]{vanLoon2005}
{van Loon}, J.~T., {Cioni}, M., {Zijlstra}, A.~A., \& {Loup}, C.
  2005{\natexlab{a}}, \aap, 438, 273

\bibitem[{{van Loon} {et~al.}(1998){van Loon}, {Hekkert}, {Bujarrabal},
  {Zijlstra}, \& {Nyman}}]{vanLoon1998}
{van Loon}, J.~T., {Hekkert}, P.~T.~L., {Bujarrabal}, V., {Zijlstra}, A.~A., \&
  {Nyman}, L.-{\AA}. 1998, \aap, 337, 141

\bibitem[{{van Loon} {et~al.}(2005{\natexlab{b}}){van Loon}, {Marshall}, \&
  {Zijlstra}}]{vanLoon2005b}
{van Loon}, J.~T., {Marshall}, J.~R., \& {Zijlstra}, A.~A. 2005{\natexlab{b}},
  \aap, 442, 597

\bibitem[{{van Loon} {et~al.}(2010{\natexlab{a}}){van Loon}, {Oliveira},
  {Gordon}, {Sloan}, \& {Engelbracht}}]{vanLoon2010b}
{van Loon}, J.~T., {Oliveira}, J.~M., {Gordon}, K.~D., {Sloan}, G.~C., \&
  {Engelbracht}, C.~W. 2010{\natexlab{a}}, \aj, 139, 1553

\bibitem[{{van Loon} {et~al.}(1996){van Loon}, {Zijlstra}, {Bujarrabal}, \&
  {Nyman}}]{vanLoon1996}
{van Loon}, J.~T., {Zijlstra}, A.~A., {Bujarrabal}, V., \& {Nyman}, L.-{\AA}.
  1996, \aap, 306, L29

\bibitem[{{van Loon} {et~al.}(2001){van Loon}, {Zijlstra}, {Bujarrabal}, \&
  {Nyman}}]{vanLoon2001}
---. 2001, \aap, 368, 950

\bibitem[{{van Loon} {et~al.}(1997){van Loon}, {Zijlstra}, {Whitelock},
  {Waters}, {Loup}, \& {Trams}}]{vanLoon1997}
{van Loon}, J.~T., {Zijlstra}, A.~A., {Whitelock}, P.~A., {et~al.} 1997, \aap,
  325, 585

\bibitem[{{van Loon} {et~al.}(2010{\natexlab{b}}){van Loon}, {Oliveira},
  {Gordon}, {Meixner}, {Shiao}, {Boyer}, {Kemper}, {Woods}, {Tielens},
  {Marengo}, {Indebetouw}, {Sloan}, \& {Chen}}]{vanLoon2010}
{van Loon}, J.~T., {Oliveira}, J.~M., {Gordon}, K.~D., {et~al.}
  2010{\natexlab{b}}, \aj, 139, 68

\bibitem[{{Villaver} {et~al.}(2003){Villaver}, {Stanghellini}, \&
  {Shaw}}]{Villaver2003}
{Villaver}, E., {Stanghellini}, L., \& {Shaw}, R.~A. 2003, \apj, 597, 298

\bibitem[{{Volk} {et~al.}(2011){Volk}, {Hrivnak}, {Matsuura}, {Bernard-Salas},
  {Szczerba}, {Sloan}, {Kraemer}, {van Loon}, {Kemper}, {Woods}, {Zijlstra},
  {Sahai}, {Meixner}, {Gordon}, {Gruendl}, {Tielens}, {Indebetouw}, \&
  {Marengo}}]{Volk2011}
{Volk}, K., {Hrivnak}, B.~J., {Matsuura}, M., {et~al.} 2011, \apj, 735, 127

\bibitem[{{Voors} {et~al.}(1999){Voors}, {Waters}, {Morris}, {Trams}, {de
  Koter}, \& {Bouwman}}]{Voors1999}
{Voors}, R.~H.~M., {Waters}, L.~B.~F.~M., {Morris}, P.~W., {et~al.} 1999, \aap,
  341, L67

\bibitem[{{Whitney} {et~al.}(2008){Whitney}, {Sewilo}, {Indebetouw},
  {Robitaille}, {Meixner}, {Gordon}, {Meade}, {Babler}, {Harris}, {Hora},
  {Bracker}, {Povich}, {Churchwell}, {Engelbracht}, {For}, {Block}, {Misselt},
  {Vijh}, {Leitherer}, {Kawamura}, {Blum}, {Cohen}, {Fukui}, {Mizuno},
  {Mizuno}, {Srinivasan}, {Tielens}, {Volk}, {Bernard}, {Boulanger}, {Frogel},
  {Gallagher}, {Gorjian}, {Kelly}, {Latter}, {Madden}, {Kemper}, {Mould},
  {Nota}, {Oey}, {Olsen}, {Onishi}, {Paladini}, {Panagia}, {Perez-Gonzalez},
  {Reach}, {Shibai}, {Sato}, {Smith}, {Staveley-Smith}, {Ueta}, {Van Dyk},
  {Werner}, {Wolff}, \& {Zaritsky}}]{Whitney2008}
{Whitney}, B.~A., {Sewilo}, M., {Indebetouw}, R., {et~al.} 2008, \aj, 136, 18

\bibitem[{{Williams} {et~al.}(2006){Williams}, {Borkowski}, {Reynolds},
  {Blair}, {Ghavamian}, {Hendrick}, {Long}, {Points}, {Raymond}, {Sankrit},
  {Smith}, \& {Winkler}}]{Williams2006}
{Williams}, B.~J., {Borkowski}, K.~J., {Reynolds}, S.~P., {et~al.} 2006, \apjl,
  652, L33

\bibitem[{{Wood} {et~al.}(1983){Wood}, {Bessell}, \& {Fox}}]{Wood1983}
{Wood}, P.~R., {Bessell}, M.~S., \& {Fox}, M.~W. 1983, \apj, 272, 99

\bibitem[{{Wood} {et~al.}(1986){Wood}, {Bessell}, \& {Whiteoak}}]{Wood1986}
{Wood}, P.~R., {Bessell}, M.~S., \& {Whiteoak}, J.~B. 1986, \apjl, 306, L81

\bibitem[{{Wood} {et~al.}(1987){Wood}, {Meatheringham}, {Dopita}, \&
  {Morgan}}]{Wood1987}
{Wood}, P.~R., {Meatheringham}, S.~J., {Dopita}, M.~A., \& {Morgan}, D.~H.
  1987, \apj, 320, 178

\bibitem[{{Wood} {et~al.}(1992){Wood}, {Whiteoak}, {Hughes}, {Bessell},
  {Gardner}, \& {Hyland}}]{Wood1992}
{Wood}, P.~R., {Whiteoak}, J.~B., {Hughes}, S.~M.~G., {et~al.} 1992, \apj, 397,
  552

\bibitem[{{Wooden} {et~al.}(1993){Wooden}, {Rank}, {Bregman}, {Witteborn},
  {Tielens}, {Cohen}, {Pinto}, \& {Axelrod}}]{Wooden1993}
{Wooden}, D.~H., {Rank}, D.~M., {Bregman}, J.~D., {et~al.} 1993, \apjs, 88, 477

\bibitem[{{Woods} {et~al.}(2011){Woods}, {Oliveira}, {Kemper}, {van Loon},
  {Sargent}, {Matsuura}, {Szczerba}, {Volk}, {Zijlstra}, {Sloan}, {Lagadec},
  {McDonald}, {Jones}, {Gorjian}, {Kraemer}, {Gielen}, {Meixner}, {Blum},
  {Sewi{\l}o}, {Riebel}, {Shiao}, {Chen}, {Boyer}, {Indebetouw}, {Antoniou},
  {Bernard}, {Cohen}, {Dijkstra}, {Galametz}, {Galliano}, {Gordon}, {Harris},
  {Hony}, {Hora}, {Kawamura}, {Lawton}, {Leisenring}, {Madden}, {Marengo},
  {McGuire}, {Mulia}, {O'Halloran}, {Olsen}, {Paladini}, {Paradis}, {Reach},
  {Rubin}, {Sandstrom}, {Soszy{\'n}ski}, {Speck}, {Srinivasan}, {Tielens}, {van
  Aarle}, {van Dyk}, {van Winckel}, {Vijh}, {Whitney}, \&
  {Wilkins}}]{Woods2011}
{Woods}, P.~M., {Oliveira}, J.~M., {Kemper}, F., {et~al.} 2011, \mnras, 411,
  1597

\bibitem[{{Wright} {et~al.}(2010){Wright}, {Eisenhardt}, {Mainzer}, {Ressler},
  {Cutri}, {Jarrett}, {Kirkpatrick}, {Padgett}, {McMillan}, {Skrutskie},
  {Stanford}, {Cohen}, {Walker}, {Mather}, {Leisawitz}, {Gautier}, {McLean},
  {Benford}, {Lonsdale}, {Blain}, {Mendez}, {Irace}, {Duval}, {Liu}, {Royer},
  {Heinrichsen}, {Howard}, {Shannon}, {Kendall}, {Walsh}, {Larsen}, {Cardon},
  {Schick}, {Schwalm}, {Abid}, {Fabinsky}, {Naes}, \& {Tsai}}]{Wright2010}
{Wright}, E.~L., {Eisenhardt}, P.~R.~M., {Mainzer}, A.~K., {et~al.} 2010, \aj,
  140, 1868

\bibitem[{{Zaritsky} {et~al.}(1997){Zaritsky}, {Harris}, \&
  {Thompson}}]{Zaritsky1997}
{Zaritsky}, D., {Harris}, J., \& {Thompson}, I. 1997, \aj, 114, 1002

\bibitem[{{Zickgraf} {et~al.}(1986){Zickgraf}, {Wolf}, {Leitherer},
  {Appenzeller}, \& {Stahl}}]{Zickgraf1986}
{Zickgraf}, F.-J., {Wolf}, B., {Leitherer}, C., {Appenzeller}, I., \& {Stahl},
  O. 1986, \aap, 163, 119

\bibitem[{{Zickgraf} {et~al.}(1985){Zickgraf}, {Wolf}, {Stahl}, {Leitherer}, \&
  {Klare}}]{Zickgraf1985}
{Zickgraf}, F.-J., {Wolf}, B., {Stahl}, O., {Leitherer}, C., \& {Klare}, G.
  1985, \aap, 143, 421

\end{thebibliography}

\end{document}